%% file: main.tex
\documentclass[12pt,a4paper,titlepage]{article}
\usepackage{tikz}
\usetikzlibrary{positioning, arrows.meta}
\usetikzlibrary{arrows.meta, calc}
\usetikzlibrary{positioning,fit}

\usepackage{t1enc}
\usepackage{amsmath}
\usepackage{latexsym}
\usepackage{natbib}
\usepackage{epsfig}
\usepackage{textcomp}
\usepackage{alltt}
\usepackage{graphicx}
\usepackage{lscape}
 \usepackage{hyperref}
 
\usepackage{multirow}
\usepackage{numprint}
 \usepackage{xcolor}
  
\usepackage{rotating}
\usepackage{pdflscape}

\usepackage{setspace}
\usepackage{booktabs}

\newcommand{\X}{\boldsymbol{X}}

\usepackage[nolists, tablesfirst]{endfloat}

\newcommand{\blind}{0}

\begin{document}

\setlength{\abovedisplayskip}{5 pt}
\setlength{\belowdisplayskip}{5 pt}

\newcommand{\bi}{\begin{itemize}}
\newcommand{\ei}{\end{itemize}}
\newcommand{\nothere}[1]{}
\newcommand{\noi}{\noindent}
\newcommand{\mbf}[1]{\mbox{\boldmath $#1$}}
\newcommand{\cond}{\, |\,}
\newcommand{\hO}[2]{{\cal O}_{#1}^{#2}}
\newcommand{\hF}[2]{{\cal F}_{#1}^{#2}}

\newcommand{\tl}[1]{\tilde{\lambda}_{#1}^T}
\newcommand{\la}[2]{\lambda_{#1}^T(Z^{#2})}
\newcommand{\I}[1]{1_{(#1)}}
\newcommand{\cd}{\mbox{$\stackrel{\mbox{\tiny{\cal D}}}{\rightarrow}$}}
\newcommand{\cp}{\mbox{$\stackrel{\mbox{\tiny{p}}}{\rightarrow}$}}
\newcommand{\cas}{\mbox{$\stackrel{\mbox{\tiny{a.s.}}}{\rightarrow}$}}
\newcommand{\ld}{\mbox{$\; \stackrel{\mbox{\tiny{def}}}{=3D} \; $}}
\newcommand{\nk}{\mbox{$n \rightarrow \infty$}}
\newcommand{\con}{\mbox{$\rightarrow $}}
\newcommand{\dprime}{\mbox{$\prime \vspace{-1 mm} \prime$}}
\newcommand{\Borel}{\mbox{${\cal B}$}}
\newcommand{\bevis}{\mbox{$\underline{\em{Proof}}$}}
\newcommand{\Rd}[1]{\mbox{${\Re^{#1}}$}}
\newcommand{\il}[1]{{\int_{0}^{#1}}}
\newcommand{\pl}[1]{\mbox{\bf {\LARGE #1}}}

\newtheorem{thm}{Theorem}
\newtheorem{lemma}{Lemma}
\newtheorem{prop}{Proposition}

{\Large{\bf The Ghosh-Lin and Fine-Gray models for a mix of administrative and random censoring}}
\vspace{0 cm}

\if0\blind
{
{ \large
\begin{center}
Thomas H. Scheike\\
Section of Biostatistics, Department of Public Health,\\ 
University of Copenhagen \\
\O ster farimagsgade 5, dk-1014 Copenhagen, Denmark\\
ts@biostat.ku.dk\\ \vspace{4mm}
Christian Mirian\\
Department of Otorhinolaryngology, Head and Neck Surgery,  \\
G\o dstrup Hospital, Herning, Denmark\\
\vspace{4mm}
Isao Yokota\\
Department of Biostatistics, Hokkaido University\\
Hokkaido, Japan \\ \vspace{4mm}
Giuliana Cortese  \\
Department of Statistical Sciences, University of Padova \\
Padova, Italy \\ \vspace{4mm}
\end{center}
}
} \fi

\if1\blind
{
} \fi 

\bigskip
\centerline{\sc Summary}
Recurrent events or competing risks regression models are often 
applied 
in the bio-medical setting and both can be considered as 
marginal models. In presence of right-censoring, such models
need to be adjusted to give consistent estimators. 
When censoring is administrative, marginal regression models are particularly easy to estimate. However, when censoring is
instead acting randomly,  inverse probability of censoring weighting (IPCW) adjustments are typically considered to obtain parameter estimates. 
This technique relies on a censoring-weights adjustment via 
a correct censoring model, but for administrative 
censoring the adjustment is done correctly simply by modifying the risk-set. 
In practice for large central registries or some clinical trials, 
the administrative censoring time will be known for all subjects, but 
there will typically also be a proportion of subjects that are censored at random. 
In this work, we consider two frequently used regression approaches, the Ghosh-Lin model for recurrent events with terminal events and the Fine-Gray model for competing events. For these two settings, when both administrative and random censoring are present, we demonstrate how to obtain correct estimation by dealing with the combination of the two different types of censoring relying on a minimum of modeling assumptions. 
\bigskip

\noi
\vspace{3mm}

\noi
{\it Some key words}: Administrative censoring; Competing risks, 
Marginal models; Random censoring; Recurrent events; 

\section{Introduction}

Marginal regression models for recurrent events and competing risks data are now
central tools in biomedical research. For recurrent events, marginal mean and
rate-based formulations provide a direct population-averaged interpretation of
treatment effects \cite{cook-lawless-1997,ghosh-lin-2000,cook-lawless-2007}.
Similarly, in competing risks settings, regression modeling of the cumulative
incidence function through the proportional subdistribution hazards model of
\cite{fine:gray:1999} has become standard practice. Both approaches yield
marginal effect parameters that are particularly attractive in clinical and
epidemiological studies, where interpretation at the population level is often
of primary interest.

In practice, event time data are almost invariably subject to right-censoring.
For consistent estimation of marginal regression parameters, appropriate
handling of censoring is crucial. When right censoring is purely administrative — that
is, determined by a common study end date or registry extraction date —
estimation is straightforward. Under independent censoring, administrative
censoring is fully observed and can be handled by appropriate risk-set
adjustment without further modeling assumptions. This structure is typical in
large-scale registries and many randomized clinical trials with fixed follow-up.

In contrast, when censoring occurs randomly over time due to loss to follow-up,
withdrawal, or other subject-specific mechanisms, valid inference generally
requires inverse probability of censoring weighting (IPCW). The IPCW approach
reweights observed contributions by the inverse of the estimated conditional
probability of remaining uncensored, thereby restoring representativeness of the
observed risk sets. This technique has been widely used in both recurrent event
settings \cite{ghosh-lin-2000,ghosh-lin-2002} and competing risks regression
\cite{fine:gray:1999}. However, IPCW relies on correct specification or
consistent estimation of the censoring distribution. Misspecification of the
censoring model may lead to biased estimation.

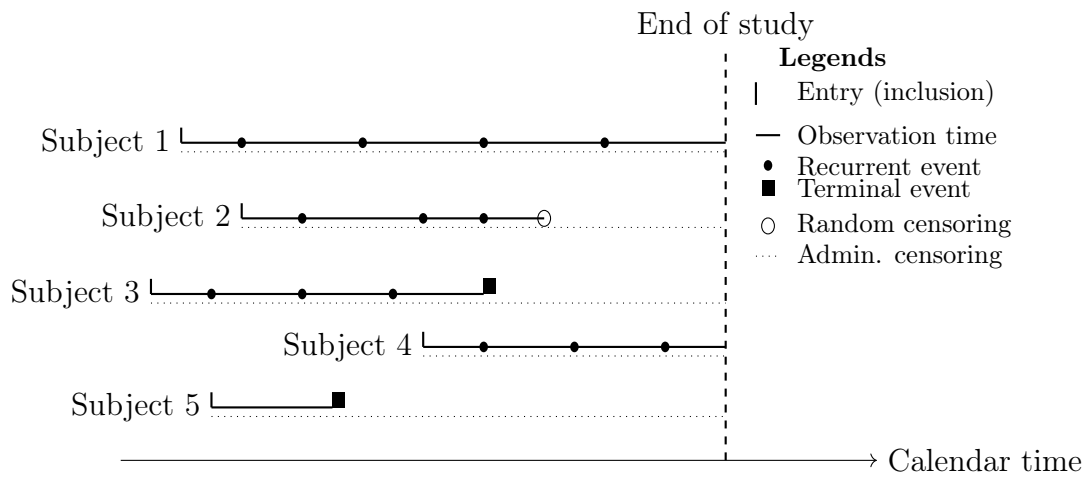
\begin{figure}
\begin{center}

\begin{tikzpicture}[xscale=0.8, yscale=1]

\draw[dashed, thick] (10,-1.2) -- (10,4.2) node[above] {End of study};

\def\offset{-0.12}

\draw[thick] (1,3) -- (10,3);
\draw[dotted] (1,3+\offset) -- (10,3+\offset);
\draw[thick] (1,3) -- +(0,0.2);
\foreach \x in {2,4,6,8}
    \fill (\x,3) circle (2pt);
\node[left] at (1,3) {Subject 1};

\draw[thick] (2,2) -- (7,2);
\draw[dotted] (2,2+\offset) -- (10,2+\offset);
\draw[thick] (2,2) -- +(0,0.2);
\foreach \x in {3,5,6}
    \fill (\x,2) circle (2pt);
\draw (7,2) circle (3pt);
\node[left] at (2,2) {Subject 2};

\draw[thick] (0.5,1) -- (6,1);
\draw[dotted] (0.5,1+\offset) -- (10,1+\offset);
\draw[thick] (0.5,1) -- +(0,0.2);
\foreach \x in {1.5,3,4.5}
    \fill (\x,1) circle (2pt);
\fill (6,1) rectangle +(0.2,0.2);
\node[left] at (0.5,1) {Subject 3};

\draw[thick] (5,0.3) -- (10,0.3);
\draw[dotted] (5,0.3+\offset) -- (10,0.3+\offset);
\draw[thick] (5,0.3) -- +(0,0.2);
\foreach \x in {6,7.5,9}
    \fill (\x,0.3) circle (2pt);
\node[left] at (5,0.3) {Subject 4};

\draw[thick] (1.5,-0.5) -- (3.5,-0.5);
\draw[dotted] (1.5,-0.5+\offset) -- (10,-0.5+\offset);
\draw[thick] (1.5,-0.5) -- +(0,0.2);
\fill (3.5,-0.5) rectangle +(0.2,0.2);
\node[left] at (1.5,-0.5) {Subject 5};

\draw[->] (0, -1.2) -- (12.5, -1.2) node[right] {Calendar time};


\begin{scope}[shift={(10.5,2.5)}, font=\footnotesize]
\node[right] at (0.20,1.6) {\textbf{Legends}};
\draw[thick] (0,1.0) -- (0,1.3);
\node[right] at (0.5,1.15) {Entry (inclusion)};
\draw[thick] (0,0.6) -- (0.4,0.6); 
\node[right] at (0.5,0.6) {Observation time};
\fill (0.2,0.2) circle (2pt);
\node[right] at (0.5,0.2) {Recurrent event};
\fill (0.1,-0.2) rectangle +(0.2,0.2);
\node[right] at (0.5,-0.1) {Terminal event};
\draw (0.2,-0.6) circle (3pt);
\node[right] at (0.5,-0.6) {Random censoring};
\draw[dotted] (0,-1.0) -- (0.4,-1.0);
\node[right] at (0.5,-1.0) {Admin. censoring};
\end{scope}
\end{tikzpicture}
\end{center}
\caption{Illustration of observation scheme for recurrent events setting with administrative and random censoring in a typical registry study taking place in calendar time.}
\label{fig:censoring}
\end{figure}

Modern biomedical data sources increasingly combine features of both censoring
types. In many national registries and pragmatic clinical trials, the
administrative censoring time is known and common across subjects, yet a
non-negligible proportion of individuals may also experience additional random
censoring prior to study termination. Thus, the observed censoring mechanism is
a mixture of deterministic administrative censoring and stochastic
subject-specific censoring, see Figure \ref{fig:censoring}. 
Despite the ubiquity of this situation,
methodological discussions typically treat censoring as either fully
administrative or fully random.

The purpose of this paper is to develop a unified framework for marginal
regression analysis of recurrent events and competing risks data under combined
administrative and random censoring. We demonstrate how the known administrative
censoring structure can be exploited to reduce reliance on modeling assumptions,
while random censoring components are accommodated through appropriately
constructed weighting schemes. Our approach requires only minimal assumptions on
the censoring mechanism and preserves the marginal interpretation of regression
parameters. The resulting estimators retain consistency and asymptotic normality
under standard regularity conditions, while relying on fewer
modeling assumptions 
compared with approaches that treat all 
censoring as purely random.

The remainder of the paper is organized as follows. Section~2 introduces
notation and the general modeling framework and subsequently modifies existing 
estimation procedures and establishes large-sample properties for our setting focussing on the 
Ghosh-Lin model.  
Section~3 briefly illustrates how the approach should be modified when considering the
Fine-Gray model. 
Section~4 presents simulation studies evaluating finite-sample performance. Section~5
illustrates the methodology using data from a clinical registry with both
administrative and random censoring. We conclude with a discussion of practical
implications and possible extensions.

\section{Notation and regression models}
\label{sect2}

We start by formulating the result for the marginal Ghosh-Lin model \citep{ghosh-lin-2002} that is a mean model for the
mean number of events seen over time, and then later in a brief remark present the small modification that
is needed for the Fine-Gray competing risks model 
\citep{fine:gray:1999}

Let $N^{*}(t)$ be the counting process of the number of recurrent events
observed over a time-period $[0,t]$, where $t\leq\tau$ for some constant
$\tau$, and $D$ denote the survival time to a terminal event. If a subject has
the terminal event,  he/she can not experience further recurrent events after
$D$. Thus we only observe the recurrent event processes up to $\tau \wedge D$,
where $a \wedge b = \min(a,b)$, such that  $N^*(t)=N^{*}(t \wedge D)$ because
subjects will only have recurrent events when still alive.

We are interested in studying the effect of a $p$-dimensional covariate vector
$X$ measured at baseline on the marginal mean number of recurrent events over time via the regression model
\begin{align}
	E(N^*(t) | X)  & = \mu_0(t) \exp(X^T\beta ) , \label{eqr:1}
\end{align}
where $\mu_0(t)$ is an unspecified baseline function that represents the marginal mean number of recurrent events up to $t$ for $X=0$, with $d\mu_0(t) = E(dN^*(t)|X=0)$, and $\beta$ quantifies the covariates effect. 
The proportional means model in \eqref{eqr:1} 
has the advantage that it does not require specifying any assumptions on the
dependence between recurrent events, and the dependence of the terminal event on recurrent events,
and has the scope to quantify the covariate effects marginally. 
Unlike Poisson processes, such models are very general and flexible, and accommodate for
possible heterogeneity between subjects, who can experience from no events to
many recurrent events, and also a terminal event. We do not assume any
dependence structures among recurrent events, but the scope is here to recover
such information directly from the history of the recurrent event process over
time using a dynamic increment estimation based on semiparametric efficiency
theory for missing data. 

In the most general case, it is enough to assume that both the recurrent events
process and the survival distribution of $D$ depend on covariates, and thus
obtain an indirect regression model for the marginal mean: 
$$ \mu(t |X)=E(N^*(t) |X)=\int_0^t S(s |X) d R(s |X), $$ where $S(t|X)=P(D> t |X)$ and the conditional rate $dR(t) = E(dN^*(t) | D> t, X)$ 
describes the recurrent event rate among the survivors. Note that, since no
further recurrent events can be experienced after the terminal event, then for
$t> D$ it is $N^*(t) = N^*(D)$,  and thus $E\{dN^*(t) |X \} = E\{dN^*(t \wedge
D)|X ) = E \{ dN^* (t) I(D \geq t) |X \}$. Unfortunately, this indirect modeling
approach is not able to measure the covariate effect directly on the marginal
mean of recurrent events, which is here our primary scope.

The observation of recurrent and terminal events may be precluded 
for subjects who are right censored during follow-up, and thus the 
process $N^*(t)$ is only observable up to the censoring time $C$. 
We shall here consider the special case where the censoring is 
a mix of administrative censoring $C_A$ and additional random 
censoring $C_R$. 
The actual censoring is thus $C= C_A \wedge C_R$, with the additional assumption 
that $C_A$ is always observed. 
Let us define $\delta=I(D \leq C)$,  $\delta_R =I(C_R \leq C_A)$, and $T=D \wedge C$. 
Moreover, let $N(t)=N^*(t \wedge C)$ be the observed number of
events and define the at-observation process as  $Y(t)=I(T \geq t)$. 
Thus, if neither $C_R$, or $D$ has occurred, subjects are no longer 
under observation, and are at most observed out to $C_A$.
The censoring scheme is illustrated in Figure \ref{fig:censoring}. 
When $C_A = \tau$ for all subjects, we return to the simple setting without 
administrative censoring where $C=C_R$ and subjects may be potentially at risk for a 
longer time.

We require that both censoring times are independent
of the outcomes $(N^*,D)$ given covariates, and in addition that the two censoring times are 
independent. The independence between the random censoring and administrative censoring is 
often reasonable, but can also be further relaxed by using a conditional censoring 
model for $C_R$ 
given both $X$ and $C_A$. We shall assume for simplicity of presentation that the 
censoring distribution of
$C_R$ given $X$ and $C_A$ does not depend on the covariate, and we denote this 
conditional survival distribution as $G(t)$.
We consider a random sample of size $n$ and assume that data come from the independent 
observations 
$\{N_{i}(t), T_i , \delta_i, X_i, C_{Ai} \}$ for $i=1,...,n$ and any $t \in [0,\tau]$.

\subsection{Inference and asymptotic results}

The administrative censoring, $C_A$, is observable  
for all subjects, while simple random censoring takes 
place through an additional censoring variable $C_R$.
The usual approach is to analyze data with $C=C_R\wedge C_A$ and apply the IPCW technique on the censoring variable $C$, without distinguishing between the two sources of censoring \citep{cook-lawless-2018}. In the paper, we refer to `IPCW adjustment' when applying this standard approach.

In this section, we show how to modify the usual 
IPCW estimating equations of  \cite{ghosh-lin-2002} to our 
setting where the IPCW adjustment is performed only for the censoring time  $C_R$, while $C_A$  is used to adjust the risk set of subjects under observation along the study period. This risk-set adjustment is denoted as `Adm' in the following sections of the paper.

Let $\hat G(t)$ be the Kaplan-Meier estimator for the distribution $G(t)$ of random censoring that is observed as right-censored by $D \wedge C_A$. 
In presence of the terminal event $D$, for each observation, 
either $C_{Ri}$ or $D_i$, or neither of the two, is observed, and as stated above 
$C_{Ai}$ is known for all subjects.
Accordingly, let us define the IPCW weights as 
$w_i(t)= I(C_{Ai} > t) I(C_{Ri} \geq D_i \wedge t) G(t) / G(T_i \wedge t)$. 
These weights can be shown to have mean value 
\begin{align*}
E( w_i(t) | C_{Ai} ) & = E[ E\{  w_i(t) | N_i, D_i, X_i, C_{Ai} \} | C_{Ai}] \\
  & =I(C_{Ai} > t) \, G(t) \, E\left[ E\left\{ 
  \frac{I(C_{Ri} \geq D_i \wedge t)}{G(D_i \wedge t)} |N_i, D_i, X_i, C_{Ai} \right\} | C_{Ai} \right] \\
  & = G(t) I(C_{Ai} > t)
\end{align*}
for $i=1, \ldots, n$, by the law of
conditional expectations and using the conditional independence, 
and they can be estimated by the observed counterpart
$\hat w_i(t) = I(C_{Ai} > t) I(C_{Ri} \geq D_i \wedge t) \hat G(t) / \hat G(T_i \wedge t)$.
If there is no additional random censoring, these weights will reduce to $I(C_{Ai} > t)$ since
also in this case $G(t) = 1$ for all $t$. 

Then, we estimate $\hat \beta$ as the solution to the weighted estimating equation 
\begin{align}
\label{eqr:4}
	U(\beta) = \sum_{i=1}^{n} \int_0^{C_{Ai}} \hat w_i(t) \{ X_i - \bar{X}(\beta, t) \} d N_i(t) = 0,              
\end{align}
where $\bar{X}(\beta, t) = \hat S_1(\beta,t) / \hat S_0(\beta,t)$ 
with $\hat S_k(\beta, t) = \sum_{i=1}^n  \hat w_i(t)  X_i^k \exp{(X_i^T\beta)}$, for $k=0,1,2$. 
We let $\bar{x}(t)=s_1(\beta,t)/s_0(\beta,t)$, where 
$s_k(\beta,t) = E[ G(t) I(C_A > t) X^k \exp( X^T \beta) ]$ is 
the limit in probability of $S_k(\beta,t)$. 

That this estimating equation is unbiased follows from the fact that 
\begin{align*}
 & E ( w(t) \{ X - \bar{x}(\beta, t) \} d[ N(t) - \mu_0(t) \exp(X^T \beta)] | X ) \\
	& = E ( I(C_A > t) G(t) \{ X - \bar{x}(\beta, t) \} d[ N(t) - \mu_0(t) \exp(X^T \beta)] | X )=0   
\end{align*}
if the Ghosh-Lin model holds 
and if the censoring model for $C_R$ is correct, still following the technical 
arguments of \cite{ghosh-lin-2002} and under the conditions stated there. 
Note also that if we use the conditional distribution of $C_R$ given $(C_A, X)$, denoted as  $G(t;X,C_A)$,
we will still get an unbiased estimating equation 
since $E(I(C_R \geq D \wedge t)/G(D \wedge t; X, C_A))=1$.

The properties of the estimator can be established by following the arguments 
of \cite{ghosh-lin-2002}, then achieving a consistent and asymptotically normal estimator with a variance we estimate. 
Defining 
$A =  E [ \partial U_i (\beta_0) / \partial \beta ] = E [ \int_0^{C_{Ai}} (X_i - \bar x(\beta_0,t))^{\otimes 2} G(t) e^{X_i^T \beta_0} d \mu_0(t) ] $, 
with $ \bar x(\beta,t) = s_1(\beta,t) / s_0(\beta,t)$, where $s_k(\beta,t) = E[G(t)X^{\otimes k} e^{X^T \beta } ] $ is the limit of $\hat S_k (\beta,t)$ for $k=0,1,2$,
and $ \hat A(\beta) = - n^{-1} \sum_{i=1}^n \partial U_i(\beta) / \partial \beta$.
It can be shown that $\hat \beta$ is a consistent estimator of $\beta_0$ and that the $i$-th
influence function  of $(\hat \beta - \beta_0 )$ is $A^{-1} \phi_i(X,\beta_0) $, 
where $\phi_i(X,\beta_0)   = \phi_i^R(X,\beta_0) + \phi_i^C(X,\beta_0)$ and 
\begin{align}
	\phi_i^R(X,\beta_0) = &  \int_0^{C_{Ai}} \{ X_i - \bar x(\beta_0, t) \} w_i(t) dM_i(t,X),  \label{eq1:phiR} \\
	\phi_i^{C}(X,\beta_0) = &  \int_0^{C_{Ai}} \frac{ q(t)}{S(t) G(t)} d M_i^{C_R}(t),
\label{eq1:phiC}
\end{align}
with 
\begin{align*}
	q(t) & = - E [ \int_t^{C_{Ai}}  \{X_i - \bar x(\beta_0, s) \} \ I(T_i < t)  \, w_i(s) \, dM_i(s,X) ] ,  \\
	& =  E [ (\tilde \mu_0(C_{Ai}) - \tilde \mu_0(t)) X_i \, \exp(X_i^T \beta) I (D_i < t )] \\
	& - E [ ( \Gamma(C_{Ai}) - \Gamma(t)) \, \exp(X_i^T \beta) I (D_i < t )],
\end{align*}
where $\tilde \mu_0 (t) = \int_0^t G(s) d \mu_0(s)$  and where $\Gamma (t) = \int_0^t \bar x(\beta_0, s)  G(s) d \mu_0(s)$. 
The equations above involve the basic mean-zero process $M_i(t,X) = N_i(t) - \mu_0(t) \exp(X^T \beta)$ and the censoring martingale
 $M^{C_R}_i(t)= N^{C_R}_i(t) -  \int_0^t Y_i(s) d\Lambda^{C_R} (s)$, 
based on the  counting process $N^{C_R}_i(t) = I(T_i \leq t, \delta =0, \delta_R =1)$ and cumulative hazard function $\Lambda^{C_R} (t) = - \log G(t) $. 
Still following the argument of \cite{ghosh-lin-2002}, it is shown  
that the normalized 
estimator $n^{1/2}(\hat \beta - \beta_0 )$ is asymptotically distributed as $N(0, \Sigma)$, with $\Sigma= A^{-1} \text{Var}\left\{\phi_1(X,\beta_0)\right\} A^{-1^T}$ and $\text{Var}\left\{\phi_1(X,\beta_0)\right\} = E\left\{ [\phi^R_1(X,\beta_0) + \phi^C_1(X,\beta_0) ]^{\otimes  2} \right\}$, 
where the outer product is $y^{\otimes  2}= y\, y^T$.
To estimate the influence function, we can plug-in estimates of all quantities and in 
particular we here note that 
\begin{align*}
\hat q(t) & = 
	n^{-1} \sum_i  \left[ \{ \hat \mu_0(C_{Ai}) - \hat \mu_0(t)\} \,  X_i \, \exp(X_i^T \hat \beta) \, I(D_i < t ) \, \frac{I(D_i<C_{Ri})}{\hat G(D_i)} \right] \\
& -  n^{-1} \sum_i    \left[ \{ \hat \Gamma(C_{Ai}) - \hat \Gamma(t)\} \, \exp(X_i^T \hat \beta) \, I (D_i < t ) \, \frac{I(D_i<C_{Ri})}{\hat G(D_i)} \right],
\end{align*}
which is a simple IPCW-estimator. 

The baseline mean number of recurrent events can be estimated by a Breslow estimator as
\begin{align*}
\hat \mu_0 (t) &= \sum_i \int_0^t \frac{1}{S_0( \hat  \beta,s) }  dN_i(s), 
\end{align*}
which has influence function $\phi_i^B(X,\beta_0) + \phi_i^{B,C}(X,\beta_0) + D(t) \phi_i(X,\beta_0)$ with 
\begin{align*}
\phi_i^B(X,\beta_0) = &  \int_0^{C_{Ai}} \frac{1}{s_0(\beta_0,t)} w_i(t) dM_i(t,X), \quad
\phi_i^{B,C}(X,\beta_0) = &  \int_0^{C_{Ai}} \frac{q^B(t)}{S(t) G(t)} d M_i^{C_R}(t), 
\end{align*}
where 
\begin{align*}
q^B(t) & = - E \left[ \int_t^{C_{Ai}}  \frac{1}{s_0(\beta_0,s)} \ I(T_i < t)  \, w_i(s) \, dM_i(s,X) \right] ,  \\
  & = - E \left[ \int_t^{C_{Ai}}   G(s) \frac{1}{s_0(\beta_0,s)}  \, dM_i(s,X) I(D_i < t)  \frac{I(D_i < C_{Ri})}{G(D_i)} \right]  \\
  &  = - E \left[ ( \Gamma^B(C_{Ai}) - \Gamma^B(t)) \, \exp(X_i^T \beta) I (D_i < t )\right], \\
D(t)  = &  -E \left[  \int_0^t \frac{s_1(\beta_0,s)}{s_0^2(\beta_0,s) } dN_i(s) \right],
\end{align*}
with $\Gamma^B (t) = \int_0^t \frac{1}{s_0(\beta_0,s)} G(s) d \mu_0(s)$. 
The estimator $\hat \mu_0(t)$ of the baseline marginal mean function is consistent 
and has asymptotically a  zero-mean Gaussian distribution. 
The influence functions can be estimated by simple plug-in estimators. 

\vspace{0.3 cm}

\noindent {\it {\bf Remark I}:}  
If neither the terminal event nor the right censoring is present in the data, i.e., 
the data are fully observed, the weights simplify such that $\hat w_i(t) =\tilde w_i(t) = 1$.

\vspace{0.3 cm}
\noindent {\it {\bf Remark II}:}  
We here distinguish carefully between the two types of censorings to take advantage of the fact that the administrative censoring
can be corrected for modifying the risk-set, and then no modelling is needed for this adjustment. 
It is common practice, see, for example, \cite{cook-lawless-2018} (page 122), to  handle the presence of both censoring $(C_A, C_R)$ simply using the combined right-censoring time $C=C^A \wedge C^R$, but then adjustment is needed for $P(C> t | X)$.

In a standard biomedical setting, 
$C^A$ will typically depend on calendar time, and if the outcome also depends on the calendar time, which will often be the case, we 
would need adjustment for this in our model. This is particularly so for registry studies that may have long follow-up.

\vspace{0.3 cm}
\noindent {\it {\bf Remark III}:}  
Inference related to the estimating equation \eqref{eqr:4} can be extended
to allow the censoring distribution to depend on covariates via strata 
$L(X) \in \{ 1, \ldots,j,\ldots K\}$, defined from the covariates $X$, such that 
$G(t,L(X)) = P( C_R > t | L(X))$. 
The modified weights are
$w_i(t,L(X)) = I(C_{Ai} >t) G(t,L(X_i)) I(C_{Ri} > D_i \wedge t) / G(D_i \wedge t, L(X_i))$, where the censoring distribution can be estimated by the
stratified Kaplan-Meier estimator, and the influence functions result in
 \begin{align*}
	 \phi_i^R(X,\beta_0) = &  \int_0^{C_{Ai}} \{ X_i - \bar x(\beta_0, t) \} w_i(t,L(X)) dM_i(t,X),  \\
	 \phi_i^{C}(X,\beta_0) = &  \int_0^{C_{Ai}} \frac{ q(t,L(X))}{S(t,L(X)) G(t,L(X))} d M_i^{C_R}(t,L(X)),
\end{align*}
where $S(t,L(X)) = P( D > t | L(X))$,
and with
\begin{align*}
q(t,j) & = E [ (\tilde \mu_0(C_{Ai},j) - \tilde \mu_0(t,j)) X_i \, \exp(X_i^T \beta) I (D_i < t ) | L(X_i)=j] \\
& - E [ ( \Gamma(C_{Ai},j) - \Gamma(t,j)) \, \exp(X_i^T \beta) I (D_i < t) | L(X_i)=j],
\end{align*}
where $\tilde \mu_0 (t,j) = \int_0^t G(s,j) d \mu_0(s)$  and $\Gamma (t,j) = \int_0^t \bar x(\beta_0, s)  G(s,j) d \mu_0(s)$. 
The Breslow estimator $\hat \mu_0(t)$ of the baseline mean number of recurrent events is also modified similarly.

\section{Fine-Gray model}
For competing risks data where events are associated to $k=1,\ldots, K$ causes, when the interest lies in modeling the cumulative incidence function, the Fine-Gray model \citep{fine:gray:1999} can be considered. 
The regression formulation  
assumes that the cumulative incidence function follows the model
 \begin{align*}
	 F_1(t,\X) = &   P( D \leq t, \epsilon=1) = 1- \exp( - \Lambda_0(t) \exp( X^T \beta))
\end{align*}
where $\Lambda_0(t)$ denotes the unspecified baseline cumulative hazard and $\beta$ the 
regression coefficients. 
The above model arises from assuming a proportional hazards structure on the subdistribution hazard function, namely
$ \lambda_1^*(t|X) =  \lambda_0^*(t) \exp{(\beta^TX)}$,
with $\Lambda_0(t) = \int_0^t \lambda_0^*(s) ds$, but inference can be carried out without explicitly working with this quantity.

Similarly to the setting of recurrent events, it is  still true that 
$$ F_1(t,X) = E(N_1(t) |X)=\int_0^t S(s |X) d R_1(s |X), $$ 
where  $N_1(t) = I(D \leq t, \epsilon=1)$ is the process that counts cause 1 events, $S(t|X)=P(D > t |X)$ is the survival function, and $dR_1(t) = E(dN_1(t) | D \geq t, X)$ is  the conditional rate for cause 1.

When combining administrative and random censoring as
in the previous section, the weights $w(t)$ need to be replaced by $w_1(t)= w(t) Y_1(t)$, where $Y_1(t) = 1-N_1(t-)$, since subjects remain in the risk set for cause 1 at $t$ even if they have experienced events from causes other than 1 before $t$. Given the estimated weights $\hat w_1(t)= \hat w(t) Y_1(t)$, we can estimate the parameters of the  Fine-Gray regression model by solving an estimating equation very similar to equation \eqref{eqr:4}:
\begin{align}
\label{eqr:5}
	U_1(\beta) = \sum_{i=1}^{n} \int_0^{C_{Ai}} \hat w_{1i}(t) \{ X_i - \bar{X}(\beta, t) \} d N_{1i}(t) = 0,              
\end{align}
where $w(t)$  is replaced by $w_1(t)$ also in $\bar{X}(\beta, t)$.
This estimating equation is also unbiased, provided the Fine-Gray model is correct, and consistency and asymptotic normality follow from  \citet{fine:gray:1999} in a straightforward manner, since the classical censoring variable $C$ is replaced by the random censoring $C_R$, while keeping the administrative censoring $C_A$ out of the risk set. The influence functions related to equation \eqref{eqr:5} now involve the martingale process $M_{1i}(t,X) = N_{1i}(t) - \int_0^t Y_{1i}(s) \lambda_0^*(s) \exp(X_i^T \beta) ds$, while the censoring martingale has the same formulation as in the recurrent events setting.

\section{Simulations}

\subsection{Recurrent events: Ghosh-Lin model}

\begin{figure}
\begin{center}
\includegraphics[width=6in]{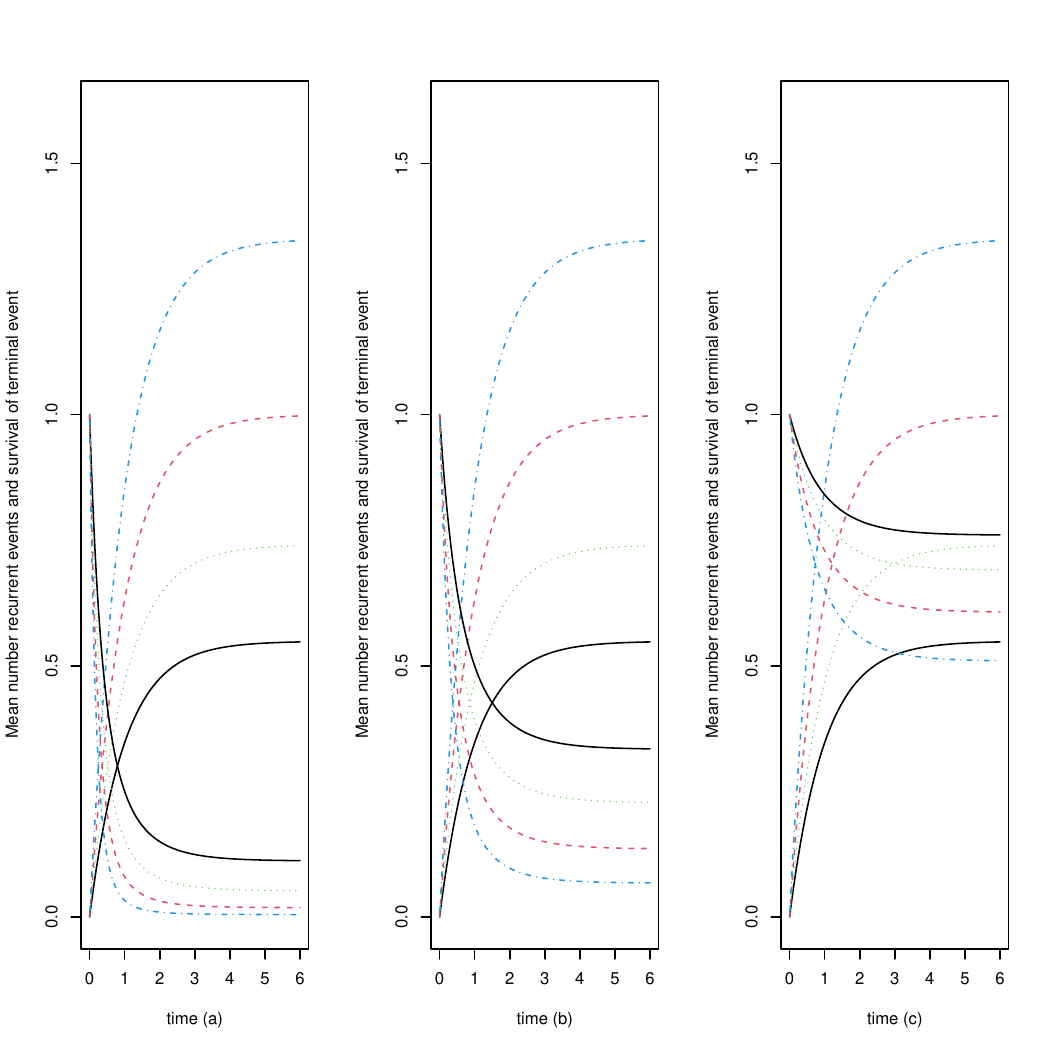}
\end{center}
	\caption{Marginal mean for recurrent events and survival function. Moderate level of mean number of recurrent events: $\rho_1=1$. Terminal event rate: (a) high level $\rho_2=4$, (b) moderate level $\rho_2=2$, (c) low level $\rho_2=0.5$.
    Black lines: $X_1=-1, X_2=1$, green lines: $X_1=-1, X_2=0$, red lines: $X_1=1, X_2=0$, blue lines: $X_1=1, X_2=1$.}
	\label{sim-ghosh-lin}
\end{figure}

We consider two independent binary covariates $\X=(X_1,X_2)$ where
$P(X_1=-1)=P(X_1=1)=0.5$ and  $P(X_2=0)=P(X_2=1)=0.5$. The marginal mean was
assumed to be $\mu_0(t) \exp( X_1 \beta_1 + X_2 \beta_2)$ with $\mu_0(t)=
[\rho_1 (1-e^{-t})]$ and $\beta_1=0.3$ and  $\beta_2=-0.3$, and the terminal
event had cumulative hazard on Cox form $\mu_1(t) \exp( -X_1 \beta_1 - X_2
\beta_2)$ with baseline $\mu_1(t)=  [\rho_2 (1-e^{-t})]$. 
To generate the conditionally independent censoring times, 
$C_A$ and $C_R$, we used the hazard function
$\lambda_c(t,X_1,X_2)=0.5 \exp( X_1 \, 0.5 - X_2 \, 0.5)$.
We refer to this as independent censoring when using the hazard function $\lambda_c(t,0,0)$
and as (covariate) dependent censoring when using the 
hazard function  $\lambda_c(t,X_1,X_2)$.
We considered different parameters for $\rho_1$ 
to control the level of the marginal mean (here we only 
report the case with 
$\rho_1=1$) and different levels of the  terminal event
($\rho_2=4, 2, 0.5$),
that provided the marginal mean and survival functions seen
in Figure \ref{sim-ghosh-lin}.
The adjustment with risk-set for administrative censoring or with standard IPCW
is only relevant for those subjects that experience the terminal event; hence, when the terminal event is seen rarely, no bias will be seen. We only report the results for sample
size $n=800$. For level $\rho_2=4$, the terminal 
event is often observed, 
making the censoring adjustment particularly important, and we stress that 
this is a difficult case where the asymptotic properties have not set in properly yet. 
We performed simulations for larger sample sizes and observed improved coverage of the 95\% confidence interval (not reported).
The more moderate levels with 
$\rho_2=2$ and $\rho_2=0.5$ showed a similar but smaller bias, and, in general, the bias decreases as the amount of terminal events decreases.

Given the marginal mean on Ghosh-Lin form and the terminal 
event on Cox form, then we simulated data using the two-stage random effects model described in \cite{Scheike2025}, with an underlying Gamma random effect with variance $1$, to
generate dependence between the recurrent events and the terminal event.
Simulation results were similar in the case of independence and for stronger dependencies (the latter not shown).
We performed simulations based on $2000$ replications, and reported the mean of estimates (Mean), bias of estimates (Bias), empirical standard error (EmpSD), mean of estimated standard
errors (MeanSE), and coverage of 95\% confidence intervals (Coverage).

We applied different methods for handling administrative and random censoring in the simulated data, in combination with the IPCW adjustment (IPCW) and/or risk-set adjustment (Adm). The different estimators that we compare are based on: \\
- data with only administrative censoring, handled with IPCW adjustment (A-IPCW); \\
- data with combined right censoring $C= C_A \wedge C_R$, handled with IPCW (RA-IPCW); \\ 
- data with only administrative censoring, handled with risk-set adjustment (Adm);\\
- data with both random censoring, handled with IPCW weighting, and administrative censoring, handled with risk-set adjustment (R-IPCW-Adm);\\
- data with both random censoring, handled with stratified IPCW
weighting, and administrative censoring, handled with risk-set adjustment (R-S-IPCW-Adm).\\
For all these cases, we investigated the regression parameters $\beta_1$ and $\beta_2$ and the baseline
estimator $\mu_0(t)$ evaluated at time-points $t=1,3$. 
The primary interest is to investigate the possible bias of the estimated parameters when censoring is not adjusted for properly. 

In the case of only administrative censoring we had on average 45 \% censoring, and 
for the combined censoring setting we had on average 13\% random censoring and 35 \% administrative censoring.  
These percentages varied slightly across the different settings.

\input{GL-sim-indep.tex}
\input{GL-sim-A1.tex}
\input{GL-sim-A1-R1.tex}

\input{GL-12-sim-indep.tex}
\input{GL-12-sim-A1.tex}
\input{GL-12-sim-A1-R1.tex}

\input{GL-1Low-sim-indep.tex}
\input{GL-1Low-sim-A1.tex}
\input{GL-1Low-sim-A1-R1.tex}


\subsubsection{Results}

Across all simulation settings, the bias patterns clearly reflect whether
censoring is handled correctly and whether the IPCW model is properly specified.
The different scenarios, shown in Figure \ref{sim-ghosh-lin},  vary by the recurrent event rate ($\rho_1$) and the terminal event
rate ($\rho_2$),  where $\rho_2 = 4$
corresponds to a high terminal event rate, see Tables  
 \ref{GL-indep}, \ref{GL-depA}, \ref{GL-depAR},
$\rho_2 = 2$ provides a moderate rate, see Tables 
\ref{GL-12-indep}, \ref{GL-12-depA}, \ref{GL-12-depAR},
and $\rho_2 = 0.5$ is associated with a low rate, see Tables 
\ref{GL-1Low-indep}, \ref{GL-1Low-depA}, \ref{GL-1Low-depAR}.

When both administrative and random censoring are independent of covariates (see Tables  \ref{GL-indep}, \ref{GL-12-indep}, \ref{GL-1Low-indep}) all methods exhibit negligible bias in the regression parameters and baseline
estimates. This includes the risk-set adjusted estimator (Adm), the combined
IPCW estimator (RA-IPCW), the proposed non-stratified IPCW estimator (R-IPCW-Adm) and
stratified estimator (R-S-IPCW-Adm). Under independence, even a
non-stratified Kaplan-Meier estimator provides a valid censoring model,
explaining the similar performance across approaches.

When administrative censoring depends on covariates, clear differences emerge (see Tables  \ref{GL-depAR}, \ref{GL-12-depAR}, \ref{GL-1Low-depAR}).
Methods relying on a non-stratified Kaplan--Meier for IPCW adjustment (A-IPCW
and RA-IPCW) exhibit noticeable bias in the regression parameters. The magnitude
of this bias increases with the terminal event rate and is most pronounced when
$\rho_2 = 4$ (high terminal event setting), smaller when $\rho_2 = 2$
(moderate), and attenuated when $\rho_2 = 0.5$ (low). This pattern is expected,
as censoring adjustment is only relevant for individuals who experience the
terminal event. Bias is also generally stronger when the rate of the terminal event
is higher.

When both administrative and random censoring depend on covariates,
misspecification becomes more consequential (see Tables  \ref{GL-depA}, \ref{GL-12-depA}, \ref{GL-1Low-depA}). The RA-IPCW and R-IPCW-Adm
estimators, which rely on non-stratified IPCW weights, show clear bias in the
regression parameters. As before, the bias is largest in the high terminal event
setting ($\rho_2 = 4$), moderate for $\rho_2 = 2$, and smallest for $\rho_2 =
0.5$. 
{However, although the censoring model is misspecified, the proposed
risk-set adjustment estimator R-IPCW-Adm shows a reduced bias, i.e., about half
of the bias under the RA-IPCW method for all settings.} 
The Adm estimator
remains
unbiased when only administrative censoring is present, as expected.
In contrast, the proposed R-S-IPCW-Adm method, which uses
covariate-stratified Kaplan-Meier weights for random censoring, together with
risk-set correction for administrative censoring, remains essentially unbiased
across all scenarios.

Overall, the simulations show that bias increases with the terminal event rate, 
particularly under censoring model misspecification, whereas the proposed R-S-IPCW-Adm estimator achieves the
intended near-unbiased performance. Coverage probabilities are generally close
to the nominal 95\% level when unbiased, 
with only 
minor deviations in the most challenging 
high-event terminal event setting.

\subsection{Competing risks: Fine-Gray model}

\begin{figure}
\begin{center}
\includegraphics[width=6in]{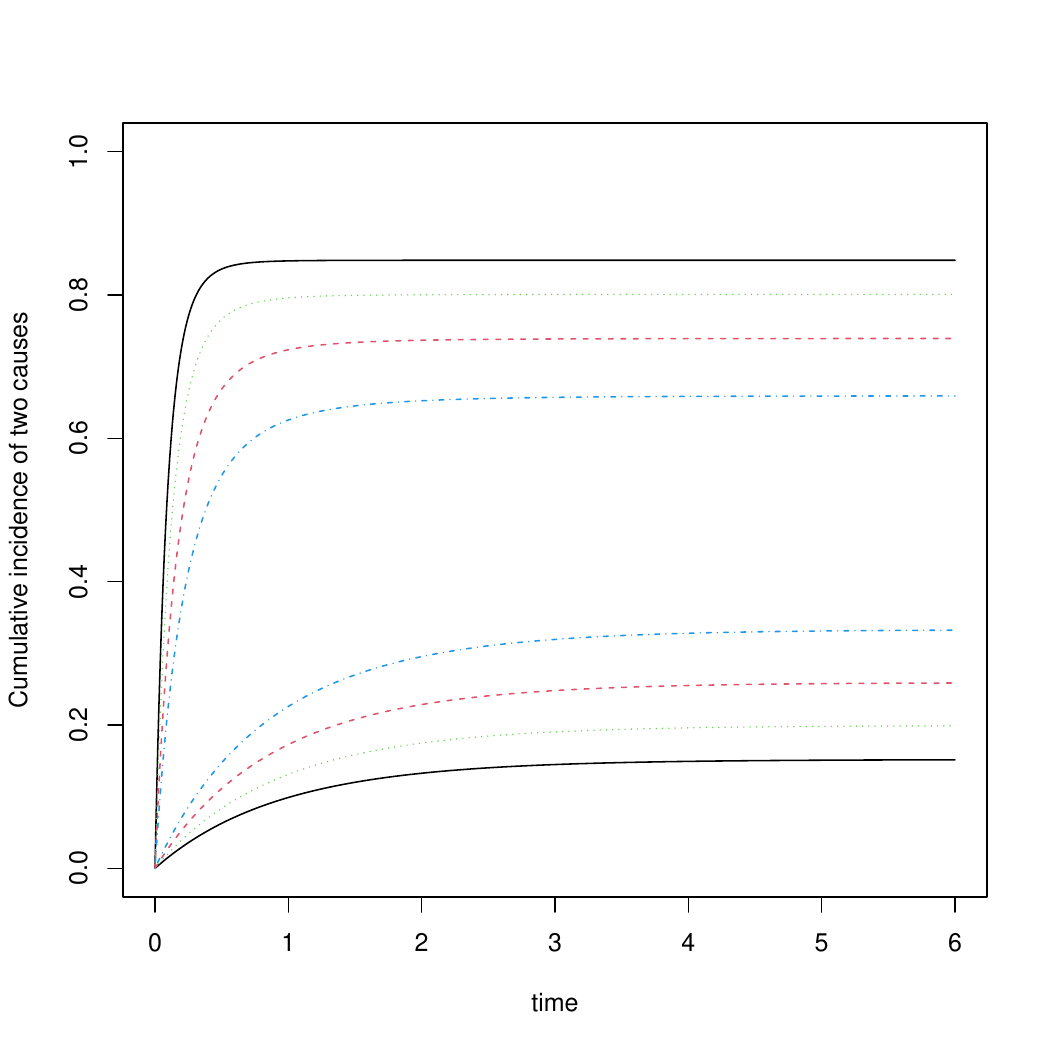}
\end{center}
\caption{Cumulative incidences for the event of interest based on the Fine-Gray model (lower curves), and for the competing event (upper curves). Setting: low cumulative
incidence of interest ($\rho_1=0.3$) and high level for the competing cause ($\rho_2=5.9$). Black lines: $X_1=-1, X_2=1$, green lines: $X_1=-1, X_2=0$, red lines: $X_1=1, X_2=0$, blue lines: $X_1=1, X_2=1$.}
\label{fig:simFG}
\end{figure}

We consider two independent binary covariates $X=(X_1,X_2)$ where
$P(X_1=-1)=P(X_1=1)=0.5$ and  $P(X_2=0)=P(X_2=1)=0.5$. 
The cumulative incidence
of cause $1$ is $F_1(t,X) = P(D \leq t, \epsilon=1 \vert X)= 1- \exp\{ \beta_0(t)
\, exp(\beta_1 X_1 + \beta_2 X_2 )\}$ with $\beta_0(t) = \log[\rho_1 (1-e^{-t})]$
and $\beta_1=0.3$ and  $\beta_2=-0.3$. The cumulative incidence of cause $2$ is
$F_2(t,X)= P(D \leq t, \epsilon=2 \vert X)$, and we assume $F_2(t,X)= \exp
\{\mu(t) \exp(-\beta_1  X_1 - \beta_2  X_2 \} \cdot\{ 1 - F_1(6,X)\}$
with $\mu(t) = \log[\rho_2 (1-e^{-t})]$. This parameterization satisfies the
constraint $F_1(t,X)+F_2(t,X) \leq 1$ for all $X=(X_1,X_2)$ and $t \in
[0,6]$, and that $F_1(t,X)$ is a Fine-Gray model. 
The survival function then is
$1-F_1(t,X)-F_2(t,X)$. 
The administrative, $C_A$, and random censoring, $C_R$,
were  generated as in the Ghosh-Lin setting.
We considered different parameters for $\rho_1$ and $\rho_2$ to control the
levels of 
cumulative incidence and here only report the case with a low cumulative
incidence of interest and a high level for the competing cause, thus having
$\rho_1=0.3$ and $\rho_2=5.9$ that gave the cumulative incidences seen in
Figure \ref{fig:simFG}. 
Adjustment of censoring is only needed when observing the competing causes. 

We performed simulations based on $2000$ replications with a sample size of $n=400$.
We report the mean of estimates (Mean), bias of estimates (Bias), empirical standard error (EmpSD), 
mean of estimated standard errors (MeanSE), and coverage of 95 \% confidence intervals (Coverage).

We considered estimators based on 
data where we
had  combined right censoring $C= C_A \wedge C_R$ handled with IPCW (RA-IPCW), 
we had only administrative censoring handled with risk-set adjustment (Adm), 
we had random censoring handled with IPCW weighting and administrative censoring 
handled with risk-set adjustment (R-IPCW-Adm), 
and finally, we had random censoring handled with stratified IPCW
weighting and administrative censoring handled with risk-set adjustment
(R-S-IPCW-Adm).
For all these cases, we investigated the regression parameters $\beta_1$ and $\beta_2$ and the baseline
estimator $\beta_0(t)$ evaluated at time-points $t=1,3$. 
The primary interest is to consider the possible bias of the estimated quantities when the 
censoring is not adjusted for properly. 

In the case of only administrative censoring we had on average 20 \% censoring, and 
for the combined censoring setting we had on average 12\% random censoring and 16 \% administrative censoring.  
These percentages varied slightly across the different settings.

\input{FG-sim-indep.tex}

\input{FG-sim-A1.tex}

\input{FG-sim-A1-R1.tex}

\subsubsection{Results}

The competing risks simulations exhibit a pattern similar to the one observed for recurrent events; see Table's \ref{FG-indep} \ref{FG-depA} \ref{FG-depAR}. 
When censoring is independent of covariates, all estimators show negligible bias and coverage
close to nominal levels.
Under covariate-dependent censoring, methods using non-stratified IPCW
adjustment (RA-IPCW and R-IPCW-Adm) display noticeable bias in the regression
parameters and baseline subdistribution quantities. This bias is more pronounced
when the competing event is frequent, as censoring adjustment is only for 
those experiencing that event.

The proposed R-S-IPCW-Adm estimator remains essentially unbiased across all
settings. Proper handling of administrative censoring through risk-set
adjustment avoids the modeling bias that arises when administrative censoring is
addressed using a misspecified IPCW model.

Overall, the simulations across both modeling frameworks demonstrate that
misspecification of the censoring model leads to systematic bias, whereas
combining risk-set correction for administrative censoring with
covariate-stratified IPCW for random censoring, as in R-S-IPCW-Adm, achieves the
intended near-unbiased performance. Coverage probabilities are generally
satisfactory and close to the nominal level with
essentially only deviations in the biased settings.

\section{Application}

We apply the mixed censoring adjustment to a standard registry study,
where as usual there are typically relatively few random 
censorings. 
Nevertheless, it is nice to have a tool to deal with this type of data 
while taking
advantage of the administrative censoring to avoid bias from incorrect
specification of the censoring weights.  Randomized clinical trials is another typical setting where there 
will be a combination of administrative and random censoring, and  a second example demonstrates 
the use of our method for such data. 

\subsection{Registry study}

In this example, 329,368 individuals were included using data from the Danish
National Patient Registry \cite{Lynge2011DanishPatientRegister}. 
All individuals aged 40.0–69.9 years who
experienced a traumatic brain injury (TBI) between 1994 and 2018 were 
included
and matched by age to five groups  of non-TBI controls. TBI severity was categorized
according to length of hospital admission (1 day, 2–3 days, or $\geq$4 days). The
cohort and its characteristics have been described in detail elsewhere 
\citep{Mirian2025,Mirian2026}.
In total, 6,935 individuals were censored due to emigration prior to the
administrative end of follow-up, and the administrative censoring was known for all subjects.

\begin{figure}
\begin{center}
\includegraphics[width=6in]{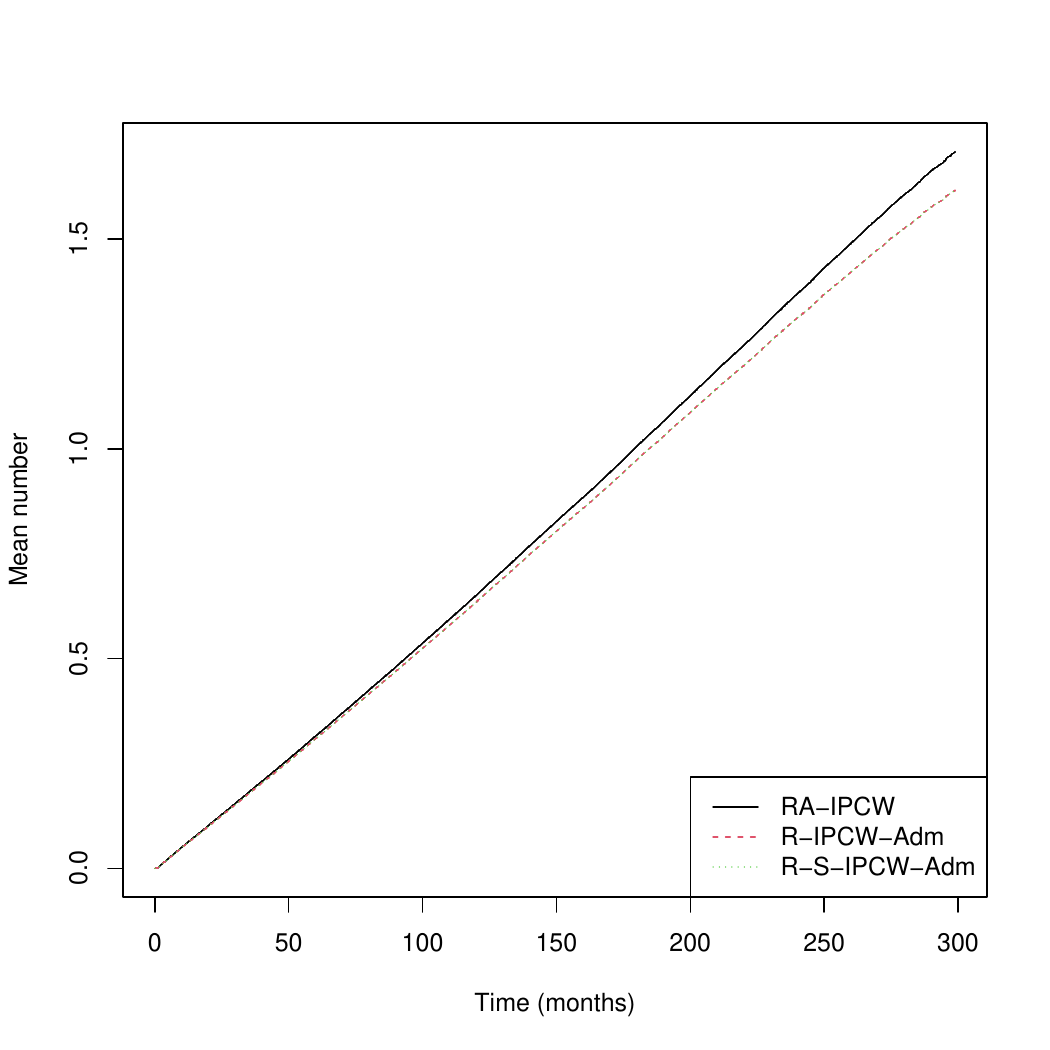}
\end{center}
\caption{Baseline mean functions for the Ghosh-Lin model based on different ways of adjusting for censoring, see text.}
\label{fig:Applighoshlin}
\end{figure}

Using the  Ghosh–Lin model (adjusted for key confounders including age)  
\citep{Mirian2025}
we investigated the number
of unique somatic comorbidities diagnosed after injury among cases and controls.
Across all models, individuals with TBI had a higher mean number of somatic
comorbidities than non-TBI controls. Comparing across RA-IPCW, R-IPCW-Adm, and
R-S-IPCW-Adm, estimates were highly consistent, with only a slight increase for
the 1-day group from RA-IPCW to the remaining two, while the other groups
remained unchanged
(Table \ref{tab:ghoshlin}).  
The mean number of comorbidities using only the baseline can be seen in Figure 
\ref{fig:Applighoshlin}, and we note that all estimates look quite similar.

\begin{figure}
\begin{center}
\includegraphics[width=6in]{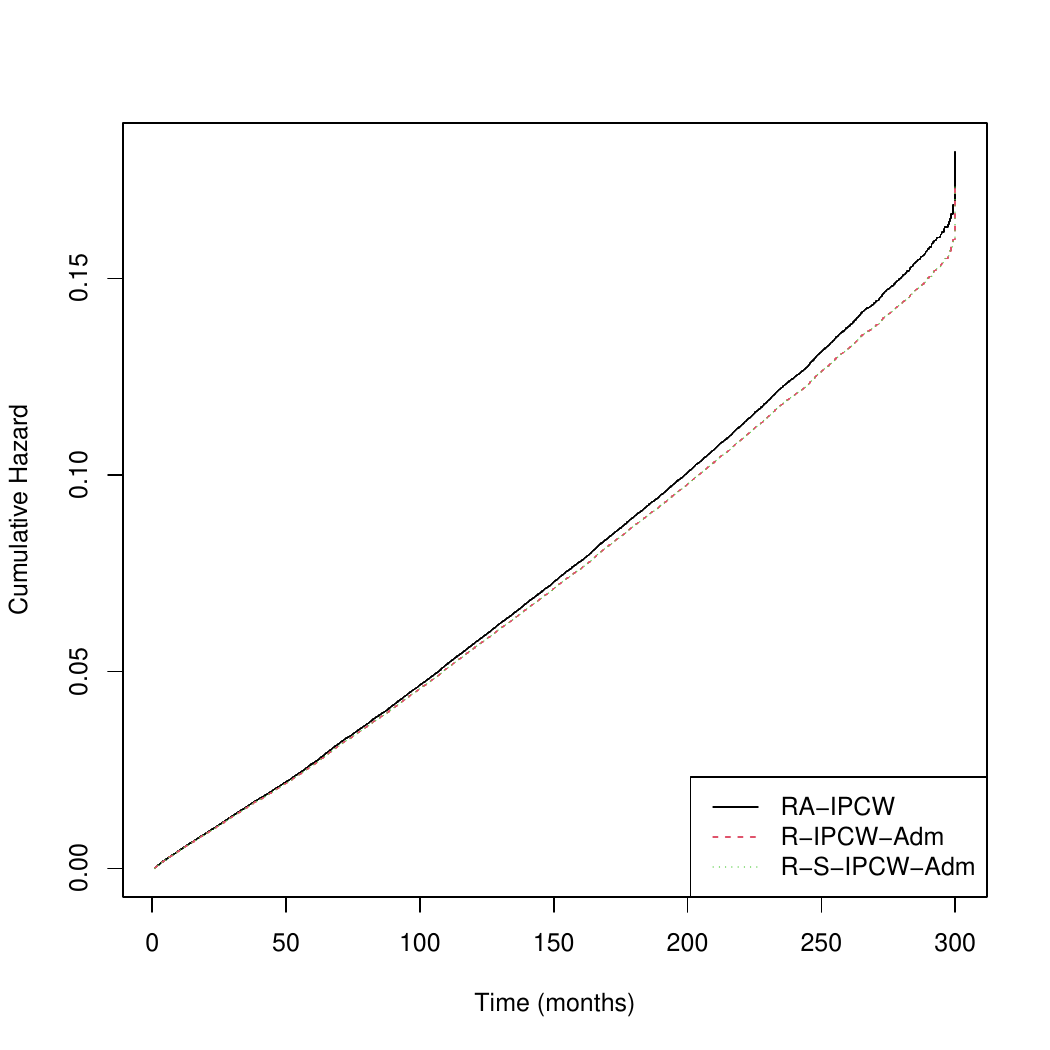}
\end{center}
\caption{Baseline subdistribution cumulative hazards for the Fine-Gray model based on different ways of adjusting for censoring, see text.}
\label{fig:Applifg}
\end{figure}

Subsequently, Fine–Gray models were adjusted
similarly to investigate the subdistribution hazard of cerebrovascular disease
post-injury (stroke, intracerebral hemorrhage)  \citep{Mirian2025}. 
Across all models, TBI was
associated with a higher hazard compared with non-TBI controls, increasing with
length of hospital admission. Across models, estimates were almost identical,
with only a slight increase for the 1-day TBI group from RA-IPCW to the
remaining two models, while the other TBI groups were unchanged 
(Table \ref{tab:finegray}).
In Figure \ref{fig:Applifg} we show the baseline of the Fine-Gray model, and again
note that all estimates are quite similar.

\begin{table}[h!]
\centering
\begin{tabular}{lccc}
\toprule
\textbf{TBI Group} & \textbf{RA-IPCW} & \textbf{R-IPCW-Adm} & \textbf{R-S-IPCW-Adm} \\
\midrule
1 day in hospital & 1.13 (1.11--1.14) & 1.15 (1.13--1.16) & 1.15 (1.13--1.16) \\
2-3 days & 1.17 (1.13--1.21) & 1.17 (1.13--1.21) & 1.17 (1.13--1.21) \\
4 or more & 1.09 (1.05--1.13) & 1.09 (1.05--1.13) & 1.09 (1.05--1.13) \\
\bottomrule
\end{tabular}
\caption{Ghosh-Lin model results: 
Ratios of Means (95\% CI) for TBI group for different censoring 
adjustment methods.
RA-IPCW: combined right censoring handled with IPCW,
R-IPCW-Adm:
random censoring handled with IPCW weighting and 
administrative censoring handled with risk-set adjustment, 
R-S-IPCW-Adm: random censoring handled with stratified
IPCW weighting and administrative censoring handled with risk-set adjustment. 
}
\label{tab:ghoshlin}
\end{table}

\begin{table}[h!]
\centering
\begin{tabular}{lccc}
\toprule
\textbf{TBI Group} & \textbf{RA-IPCW} & \textbf{R-IPCW-Adm} & \textbf{R-S-IPCW-Adm} \\
\midrule
1 day in hospital & 1.37 (1.32--1.43) & 1.40 (1.34--1.45) & 1.40 (1.34--1.45) \\
2-3 days & 1.45 (1.33--1.57) & 1.45 (1.33--1.57) & 1.45 (1.33--1.57) \\
4 or more & 1.55 (1.41--1.70) & 1.55 (1.41--1.70) & 1.55 (1.41--1.70) \\
\bottomrule
\end{tabular}
\caption{Fine-Gray model results: Subdistribution hazard ratios (95\% CI)
for TBI group for different censoring adjustment methods.
RA-IPCW: combined right censoring handled with IPCW,
R-IPCW-Adm: random censoring handled with IPCW weighting and 
administrative censoring handled with risk-set adjustment, 
R-S-IPCW-Adm: random censoring handled with stratified
IPCW weighting and administrative censoring handled with risk-set adjustment. 
}
\label{tab:finegray}
\end{table}

\subsection{RCT study}\label{sec:rct}

We further illustrate the methods using data from the NCCTG (Alliance)
intergroup trial N0147 \citep{alberts-2012},  a randomized phase III trial in
patients with resected stage III colon cancer that compared FOLFOX with or
without cetuximab. We treated recurrence as the event of interest and death
without recurrence as a competing event, which leads naturally to a Fine-Gray
regression analysis. Because the dataset records the loss to follow-up
separately from the administrative censoring, we can compare the standard IPCW
approach based on the combined censoring time with the proposed mixed-censoring
approach that handles the two components separately. After restricting to
subjects with complete baseline covariates, the analyzed sample comprised 2,496
patients, with 538 recurrences, 61 deaths without recurrence, 42 subjects lost
to follow-up, and 1,855 administratively censored observations.

We fitted the Fine -Gray model for studying  the treatment effect on recurrence
under the censoring specifications used in the simulations, together with an
additional specification, R-IPCW-Adm(arm), in which a Cox model with the
treatment arm  as predictor is assumed for the hazard of the random-censoring
event, 
with the treatment arm as the predictor so that treatment-dependent loss to
follow-up is accommodated (Table~\ref{tab:rct}). The four estimates were close.
The conventional Fine-Gray analysis gave a subdistribution hazard ratio of 1.075
and the mixed-censoring analyses gave 1.081, 
decimal place and with essentially identical standard errors. None of the
analyses gave evidence of a cetuximab effect on recurrence, in line with the
original trial report.

The administrative censoring time is latent for the subjects who recurred or
died, and for the proposed analyses we drew it from the Kaplan-Meier
administrative-censoring distribution conditional on survival beyond the
observed event time. The standard analysis does not use this time, and repeating
the imputation over ten random seeds left the proposed estimate essentially
unchanged, with a subdistribution hazard ratio between 1.07 and 1.08 throughout.
Because the trial was stopped early after interim monitoring, the observed
administrative censoring time reflects the operational end of follow-up rather
than the planned horizon, and the random censoring is close to be independent of
the treatment assignment. Separating the administrative from the random
censoring, therefore reproduces the standard analysis, and we use this example
mainly to illustrate that the proposed approach is straightforward and stable on
a real trial dataset rather than to show numerical differences between the
methods.

\begin{table}[ht]
\centering
\caption{Estimated effect of cetuximab on recurrence in trial N0147 under four different censoring specifications. Estimate (standard error) is reported on the
log-subdistribution-hazard scale, together with the subdistribution hazard ratio
(SHR) and its $95\%$ confidence interval. The latent administrative time of the
competing deaths was filled by a single Kaplan--Meier imputation.}
\label{tab:rct}
\begin{tabular}{lcc}
\hline
Method & Estimate (SE) & SHR ($95\%$ CI) \\
\hline
RA-IPCW            & $0.0722\ (0.0861)$ & $1.075\ (0.908,\ 1.273)$ \\
R-IPCW-Adm         & $0.0781\ (0.0862)$ & $1.081\ (0.913,\ 1.280)$ \\
R-S-IPCW-Adm       & $0.0783\ (0.0862)$ & $1.081\ (0.913,\ 1.280)$ \\
R-IPCW-Adm (arm)   & $0.0782\ (0.0862)$ & $1.081\ (0.913,\ 1.280)$ \\
\hline
\end{tabular}
\end{table}


\section{Discussion} 

In a setting with both administrative and random right-censoring, we have demonstrated that to avoid 
bias in model estimation when fitting marginal models, it can be useful to deal with the two types of censoring separately, and  take advantage of the administrative censoring for which 
adjustment can be done by a risk-set modification.

The separation of the two censoring types is relevant only for the marginal models considered here. It plays no role in the Cox model for a single event time, where the censoring enters only through the risk set, no censoring weights are estimated, and covariate-dependent censoring is already accommodated by conditioning on the covariates.
For the marginal models, the standard and proposed analyses agree when the administrative censoring is independent of the covariates, and differ when it depends on covariates that also affect the event process. We observed that this difference grows with the rate of the terminal or competing event, because these models retain a subject's contribution, and thus the administrative censoring time, after that event.
For the Fine-Gray model, a subject who has experienced a competing event remains in the risk set for the cause of interest, so the analysis still needs that subject's administrative censoring time; because this time is known rather than modeled, the most delicate part of subdistribution-hazard estimation needs no censoring model at all. This covariate dependence is unlikely to arise in a randomized trial, where the administrative censoring is independent of the treatment arm. It can arise in registry studies with staggered entry, where the administrative censoring time is determined by the calendar time of entry; when the case mix changes over a long inclusion period, the covariates become associated with entry time, and hence with the administrative censoring.

Taking advantage of the risk-set modification to deal with the administrative
censoring comes with a small prize in terms of efficiency. An IPCW estimator is
more efficient when the censoring weights are estimated than when they are taken
as known \citep{robins-1994}, which may seem paradoxical \citep{henmi-2004}. By
taking the administrative censoring as known, we sacrifice this gain.
{In
our simulations we observed that handling the administrative censoring with IPCW
adjustment led to a slightly improved efficiency for 
the  Fine-Gray model simulations. }
In general, unfortunately, getting the
locally efficient estimator for a marginal regression model is, however, quite
complicated \citep{ts-tm-bo-fg-2022} and requires additional 
complex modelling. 

\if0\blind
{
The methods for combined handling of random and administrative censoring have been
implemented in the {\bf mets}-package for {\bf R} and are illustrated in a vignette \citep{metsR}. 

\section*{Acknowledgement}
 
We thank Professor Therese Ovesen for allowing us to use the traumatic brain injury dataset.
This publication is based on research using information obtained from www.projectdatasphere.org, which is maintained by Project Data Sphere. Neither Project Data Sphere nor the owner(s) of any information from the web site have contributed to, approved or are in any way responsible for the contents of this publication.
} \fi

\bibliography{ghosh-lin}
\bibliographystyle{apalike}

\section*{Appendix} 

\subsection{Appendix A: Asymptotic Properties}

The basic structure of the proof follows closely the arguments of
\cite{ghosh-lin-2002} but careful attention to the our mixed
censoring pattern that is used in our the estimating equation
(2). 
 
With algebraic manipulations and using the uniform convergence 
of $\hat G(t)$ to $G(t) $, we can write the estimating equation for $\beta$  in the form given in  (2) 
 \begin{align*}
	 n^{-1/2}  U(\beta)  & =   n^{-1/2}  \sum_{i=1}^n \int_0^{C_{A_i}} \left( X_i - \bar X(\beta, t) \right) w_i(t) dM_i(t,X)    \nonumber \\
	 &  + n^{-1/2}  \sum_{i=1}^n \int_0^{C_{A_i}} \left( X_i - \bar X(\beta, t) \right) 
\left( \frac{\hat G(t)}{\hat G(T_i \wedge t)} - \frac{ G(t)}{ G(T_i \wedge t)} \right)
I(C_{Ri} \geq D_i \wedge t)  dM_i(t,X).
\end{align*}
Then, using the martingale representation of the K-M estimator,  we obtain
 \begin{align*}
	 n^{1/2}  \left( \frac{\hat G(t)}{\hat G(T_i \wedge t)} - \frac{ G(t)}{ G(T_i \wedge t)} \right) & =    - n^{-1/2} G(t) \frac{ I(T_i < t ) }{G(T_i \wedge t)} \sum_{j=1}^n \int_{T_i}^{t}  \frac{1}{Y_{\bullet}(s)} dM_j^C(s)  + o_P(1) \\
	 & =    - n^{-1/2} G(t) \frac{ I(T_i < t ) }{ G(T_i \wedge t)} \sum_{j=1}^n \int_{0}^{t}  I(s > T_i) \frac{1}{G(s) S(s)} dM_j^C(s) + o_P(1), 
 \end{align*}
 with $ Y_{\bullet}(s) = \sum_{i=1}^n Y_i(s)$. Because of the identity $ Y_{\bullet}(s) = n \hat G(s) \hat S(s)$, the last equation is obtained using the consistency of $\hat G(t)$ and $\hat S(t)$.
Then, plugging this identity into the above estimating equation  $n^{-1/2}  U(\beta) $ and changing the order of integration, we observe that
\small 
 \begin{align*}
	 & n^{-1/2}  \sum_{i=1}^n \int_0^{C_{A_i}} \left( X_i - \bar X(\beta, t) \right) \left( \frac{\hat G(t)}{\hat G(T_i \wedge t)} - \frac{ G(t)}{ G(T_i \wedge t)} \right)
   I(C_{Ri} \geq D_i \wedge t)  dM_i(t,X) \\
	 & = n^{-1/2}  \sum_{i=1}^n \int_0^{C_{A_i}} \left( X_i - \bar X(\beta, t) \right) w_i(t) I(T_i < t) \left( - n^{-1}  \sum_{j=1}^n \int_{0}^{t} I(s > T_i) \frac{1}{G(s) S(s)} dM_j^C(s)  \right) dM_i(t,X) \\
	 & = n^{-1/2}  \sum_{i=1}^n \int_0^{C_{A_i}} \left( - n^{-1}  \sum_{j=1}^n \int_{t}^{C_{A_i}}  \left( X_j - \bar X(\beta, s) \right) w_j(s) I( T_j < s) dM_j(s,X)    \right) I(T_i<t)  \frac{1}{G(t) S(t)} dM_i^C(t) \\
	 & = n^{-1/2}  \sum_{i=1}^n \int_0^{C_{A_i}} \hat q(t) \frac{1}{G(t) S(t)} dM_i^C(t)
\end{align*}
with
 \begin{align*}
	 \hat q(t)   &  = - n^{-1}  \sum_{i=1}^n \int_{t}^{C_{A_i}}  \left( X_i - \bar X(\beta, s) \right)  I(T_i < t )  w_i(s) dM_i(s,X) 
\end{align*}
 
Replacing  $\bar X(\beta, t)$ with the limit $ \bar x(\beta, t) = s_1(\beta, t) / s_0(\beta, t) $, where 
$s_k(\beta, t) = E(G(t) X^{\otimes k} \exp{(\beta^T X_i)} )$,
and it can be shown that $\hat q(t)$ has the limit in probability 
\begin{align*}
	q(t) & = - E \left[ \int_t^{C_{Ai}}  \{X_i - \bar x(\beta_0, s) \} \ I(T_i < t)  \, w_i(s) \, dM_i(s,X) \right] ,  \\
	& = - E \left[ \int_t^{C_{Ai}}  \{X_i - \bar x(\beta_0, s) \} \, G(s)  \, dM_i(s,X) \, I(D_i < t) \,  \frac{I(D_i < C_{Ri})}{G(D_i)} \right] ,\\
	& = - E \left[ \int_t^{C_{Ai}}  \{X_i - \bar x(\beta_0, s) \} \, G(s)  \, dM_i(s,X) \, I(D_i < t) \right]\\
	& =  E \left[ \int_{t}^{C_{Ai}}  \{X_i - \bar x(\beta_0, s) \} \, G(s)  \, \exp(X_i^T \beta) \, d \mu_0(s) \, I (D_i < t )\right] \\ 
	& =  E \left[ (\tilde \mu_0(C_{Ai}) - \tilde \mu_0(t)) X_i \, \exp(X_i^T \beta)\,  I (D_i < t ) \right] \\
	& - E \left[ ( \Gamma(C_{Ai}) - \Gamma(t)) \, \exp(X_i^T \beta) \, I (D_i < t )\right].
\end{align*}
On the support $\{T_i<t\le s,\, w_i(s)>0\}$, $T_i$ cannot equal $C_{Ri}$ since 
this would contradict the indicator $I(C_{Ri}\ge D_i\wedge s)$ in $w_i(s)$, 
so $I(T_i<t)$ may be replaced by $I(D_i<t)$.


\end{document}

%% file: GL-sim-indep.tex
\begin{table}[ht]
\centering
\begin{tabular}{l|l|rrrrrr}
   \hline
Method  & Par. & True & Mean & Bias & EmpSD & MeanSE & Coverage \\ 
  \hline
\multirow{4}{5.3em}{ A-IPCW} & $\beta_1$ & 0.300 & 0.296 & -0.004 & 0.097 & 0.092 & 0.932 \\ 
     & $\beta_2$ &  -0.300 & -0.298 & 0.002 & 0.177 & 0.172 & 0.939 \\ 
     & $\mu_0(1)$ & 0.632 & 0.643 & 0.011 & 0.069 & 0.070 & 0.953 \\ 
    & $\mu_0(3)$  & 0.950 & 0.964 & 0.014 & 0.112 & 0.115 & 0.947 \\ 
   \hline
\multirow{4}{5.3em}{RA-IPCW} & $\beta_1$ & 0.300 & 0.297 & -0.003 & 0.097 & 0.093 & 0.941 \\ 
   & $\beta_2$ & -0.300 & -0.303 & -0.003 & 0.187 & 0.175 & 0.929 \\ 
   & $\mu_0(1)$  & 0.632 & 0.645 & 0.012 & 0.074 & 0.073 & 0.948 \\ 
   & $\mu_0(3)$  & 0.950 & 0.964 & 0.014 & 0.125 & 0.127 & 0.944 \\ 
   \hline
\multirow{4}{5.3em}{Adm} & $\beta_1$  & 0.300 & 0.296 & -0.004 & 0.097 & 0.093 & 0.938 \\ 
     & $\beta_2$ & -0.300 & -0.299 & 0.001 & 0.178 & 0.173 & 0.942 \\ 
     &  $\mu_0(1)$ & 0.632 & 0.643 & 0.011 & 0.069 & 0.069 & 0.946 \\ 
     &  $\mu_0(3)$ & 0.950 & 0.964 & 0.014 & 0.116 & 0.113 & 0.937 \\ 
   \hline
\multirow{4}{5.3em}{R-IPCW-Adm} & $\beta_1$ & 0.300 & 0.297 & -0.003 & 0.097 & 0.094 & 0.942 \\ 
     & $\beta_2$ & -0.300 & -0.303 & -0.003 & 0.188 & 0.176 & 0.932 \\ 
   &  $\mu_0(1)$  & 0.632 & 0.645 & 0.012 & 0.075 & 0.073 & 0.943 \\ 
   &  $\mu_0(3)$  & 0.950 & 0.964 & 0.013 & 0.129 & 0.127 & 0.936 \\ 
   \hline
\multirow{4}{5.3em}{R-S-IPCW-Adm} & $\beta_1$  & 0.300 & 0.298 & -0.002 & 0.095 & 0.093 & 0.948 \\ 
   & $\beta_2$  & -0.300 & -0.305 & -0.005 & 0.183 & 0.174 & 0.939 \\ 
   &  $\mu_0(1)$  & 0.632 & 0.645 & 0.013 & 0.074 & 0.073 & 0.945 \\ 
   &  $\mu_0(3)$  & 0.950 & 0.962 & 0.012 & 0.128 & 0.127 & 0.945 \\ 
   \hline
\end{tabular}
\caption{Simulation results for Ghosh-Lin model with independent censoring. 
	Sample size n=800, $\rho_1=1$ and $\rho_2=4$. 
	A-IPCW: only administrative censoring handled with IPCW adjustment; RA-IPCW: combined right censoring handled with IPCW; Adm: only administrative censoring handled with risk-set adjustment; R-IPCW-Adm: random censoring handled with IPCW weighting, and administrative censoring handled with risk-set adjustment;  R-IPCW-S-Adm: random censoring handled with stratified IPCW weighting, and administrative censoring handled with risk-set adjustment. Mean of estimates (Mean), bias of estimates (Bias), empirical standard error (EmpSD), mean of estimated standard errors (MeanSE), and coverage of 95 \% confidence intervals (Coverage)} 
\label{GL-indep}
\end{table}

%% file: GL-sim-A1.tex
\begin{table}[ht]
\centering
\begin{tabular}{l|l|rrrrrr}
 \hline
Method & Par.  & True & Mean & Bias & EmpSD & MeanSE & Coverage \\ 
 \hline
\multirow{4}{5.3em}{ A-IPCW} & $\beta_1$  & 0.300 & 0.260 & -0.040 & 0.140 & 0.122 & 0.919 \\ 
  & $\beta_2$  & -0.300 & -0.262 & 0.038 & 0.252 & 0.226 & 0.933 \\ 
  & $\mu_0(1)$   & 0.632 & 0.646 & 0.014 & 0.090 & 0.087 & 0.943 \\ 
  & $\mu_0(3)$  & 0.950 & 0.964 & 0.014 & 0.154 & 0.147 & 0.917 \\ 
   \hline
\multirow{4}{5.3em}{RA-IPCW} & $\beta_1$  & 0.300 & 0.265 & -0.035 & 0.137 & 0.119 & 0.927 \\ 
   & $\beta_2$ & -0.300 & -0.266 & 0.034 & 0.253 & 0.222 & 0.928 \\ 
  & $\mu_0(1)$  & 0.632 & 0.648 & 0.015 & 0.095 & 0.089 & 0.940 \\ 
   & $\mu_0(3)$ & 0.950 & 0.965 & 0.015 & 0.177 & 0.163 & 0.904 \\ 
   \hline
\multirow{4}{5.3em}{Adm} & $\beta_1$  & 0.300 & 0.303 & 0.003 & 0.139 & 0.121 & 0.910 \\ 
   & $\beta_2$ & -0.300 & -0.300 & -0.000 & 0.251 & 0.227 & 0.930 \\ 
  & $\mu_0(1)$ & 0.632 & 0.645 & 0.013 & 0.090 & 0.086 & 0.940 \\ 
   & $\mu_0(3)$ & 0.950 & 0.958 & 0.008 & 0.156 & 0.144 & 0.904 \\ 
   \hline
\multirow{4}{5.3em}{R-IPCW-Adm} & $\beta_1$ & 0.300 & 0.304 & 0.004 & 0.136 & 0.119 & 0.911 \\ 
   & $\beta_2$ & -0.300 & -0.301 & -0.001 & 0.251 & 0.223 & 0.931 \\ 
  & $\mu_0(1)$  & 0.632 & 0.645 & 0.013 & 0.096 & 0.089 & 0.934 \\ 
   & $\mu_0(3)$ & 0.950 & 0.958 & 0.008 & 0.179 & 0.161 & 0.893 \\ 
   \hline
\multirow{4}{5.3em}{R-S-IPCW-Adm} & $\beta_1$ & 0.300 & 0.308 & 0.008 & 0.133 & 0.116 & 0.914 \\ 
  & $\beta_2$ & -0.300 & -0.305 & -0.005 & 0.246 & 0.219 & 0.937 \\ 
 & $\mu_0(1)$  & 0.632 & 0.646 & 0.014 & 0.094 & 0.089 & 0.941 \\ 
  & $\mu_0(3)$ & 0.950 & 0.952 & 0.002 & 0.176 & 0.159 & 0.886 \\ 
   \hline
\end{tabular}
\caption{Simulation results for Ghosh-Lin model with dependent administrative censoring and independent random censoring. 
Sample size n=800, $\rho_1=1$ and $\rho_2=4$. 
A-IPCW: only administrative censoring handled with IPCW adjustment; RA-IPCW: combined right censoring handled with IPCW; Adm: only administrative censoring handled with risk-set adjustment; R-IPCW-Adm: random censoring handled with IPCW weighting, and administrative censoring handled with risk-set adjustment;  R-IPCW-S-Adm: random censoring handled with stratified IPCW weighting, and administrative censoring handled with risk-set adjustment. Mean of estimates (Mean), bias of estimates (Bias), empirical standard error (EmpSD), mean of estimated standard errors (MeanSE), and coverage of 95 \% confidence intervals (Coverage)} 
	\label{GL-depA}
\end{table}

%% file: GL-sim-A1-R1.tex
\begin{table}[ht]
\centering
\begin{tabular}{l|l|rrrrrr}
   \hline
Method & Par.  & True & Mean & Bias & EmpSD & MeanSE & Coverage \\ 
   \hline
\multirow{4}{5.3em}{ A-IPCW} & $\beta_1$ & 0.300 & 0.270 & -0.030 & 0.070 & 0.070 & 0.933 \\ 
     & $\beta_2$  & -0.300 & -0.272 & 0.028 & 0.140 & 0.137 & 0.944 \\ 
    & $\mu_0(1)$ & 0.632 & 0.637 & 0.005 & 0.060 & 0.060 & 0.953 \\ 
   & $\mu_0(3)$ & 0.950 & 0.956 & 0.005 & 0.093 & 0.093 & 0.945 \\ 
   \hline
\multirow{4}{5.3em}{ RA-IPCW} & $\beta_1$ & 0.300 & 0.246 & -0.054 & 0.074 & 0.074 & 0.883 \\ 
    & $\beta_2$  & -0.300 & -0.251 & 0.049 & 0.150 & 0.146 & 0.935 \\ 
   & $\mu_0(1)$& 0.632 & 0.634 & 0.002 & 0.065 & 0.064 & 0.938 \\ 
  & $\mu_0(3)$  & 0.950 & 0.952 & 0.002 & 0.104 & 0.102 & 0.936 \\ 
   \hline
\multirow{4}{5.3em}{Adm} & $\beta_1$ & 0.300 & 0.300 & -0.000 & 0.071 & 0.071 & 0.955 \\ 
     & $\beta_2$  & -0.300 & -0.297 & 0.003 & 0.141 & 0.138 & 0.947 \\ 
   & $\mu_0(1)$ & 0.632 & 0.638 & 0.006 & 0.060 & 0.060 & 0.951 \\ 
   & $\mu_0(3)$  & 0.950 & 0.955 & 0.005 & 0.094 & 0.091 & 0.938 \\ 
   \hline
\multirow{4}{5.3em}{R-IPCW-Adm} & $\beta_1$ & 0.300 & 0.272 & -0.028 & 0.074 & 0.074 & 0.931 \\ 
    & $\beta_2$  & -0.300 & -0.275 & 0.025 & 0.151 & 0.147 & 0.944 \\ 
    & $\mu_0(1)$ & 0.632 & 0.637 & 0.005 & 0.066 & 0.064 & 0.942 \\ 
   & $\mu_0(3)$  & 0.950 & 0.954 & 0.004 & 0.106 & 0.103 & 0.933 \\ 
   \hline
\multirow{4}{5.3em}{R-S-IPCW-Adm} & $\beta_1$ & 0.300 & 0.299 & -0.001 & 0.073 & 0.074 & 0.948 \\ 
     & $\beta_2$  & -0.300 & -0.298 & 0.002 & 0.149 & 0.146 & 0.942 \\ 
    & $\mu_0(1)$ & 0.632 & 0.638 & 0.006 & 0.065 & 0.064 & 0.943 \\ 
   & $\mu_0(3)$  & 0.950 & 0.953 & 0.003 & 0.105 & 0.103 & 0.936 \\ 
   \hline
\end{tabular}
\caption{Simulation results for Ghosh-Lin model with dependent administrative censoring and dependent random censoring. 
	Sample size n=800, $\rho_1=1$ and $\rho_2=4$. 
	A-IPCW: only administrative censoring handled with IPCW adjustment; RA-IPCW: combined right censoring handled with IPCW; Adm: only administrative censoring handled with risk-set adjustment; R-IPCW-Adm: random censoring handled with IPCW weighting and administrative censoring handled with risk-set adjustment;  R-IPCW-S-Adm: random censoring handled with stratified IPCW weighting and administrative censoring handled with risk-set adjustment. Mean of estimates (Mean), bias of estimates (Bias), empirical standard error (EmpSD), mean of estimated standard errors (MeanSE), and coverage of 95 \% confidence intervals (Coverage)} 
\label{GL-depAR}
\end{table}

%% file: GL-12-sim-indep.tex
\begin{table}[ht]
\centering
\begin{tabular}{l|l|rrrrrr}
   \hline
Method & Par.  & True & Mean & Bias & EmpSD & MeanSE & Coverage \\ 
   \hline
\multirow{4}{5.3em}{A-IPCW} & $\beta_1$   & 0.300 & 0.296 & -0.004 & 0.074 & 0.071 & 0.947 \\ 
  & $\beta_2$   & -0.300 & -0.301 & -0.001 & 0.141 & 0.138 & 0.952 \\ 
    & $\mu_0(1)$  & 0.632 & 0.642 & 0.010 & 0.059 & 0.060 & 0.961 \\ 
    & $\mu_0(3)$  & 0.950 & 0.961 & 0.011 & 0.090 & 0.094 & 0.957 \\ 
   \hline
\multirow{4}{5.3em}{RA-IPCW} & $\beta_1$   & 0.300 & 0.295 & -0.005 & 0.077 & 0.076 & 0.951 \\ 
   & $\beta_2$ & -0.300 & -0.298 & 0.002 & 0.149 & 0.146 & 0.953 \\ 
   & $\mu_0(1)$ & 0.632 & 0.641 & 0.008 & 0.063 & 0.064 & 0.964 \\ 
    & $\mu_0(3)$  & 0.950 & 0.959 & 0.009 & 0.100 & 0.103 & 0.955 \\ 
   \hline
\multirow{4}{5.3em}{Adm} & $\beta_1$   & 0.300 & 0.296 & -0.004 & 0.075 & 0.072 & 0.949 \\ 
    & $\beta_2$ & -0.300 & -0.300 & -0.000 & 0.142 & 0.139 & 0.949 \\ 
    & $\mu_0(1)$  & 0.632 & 0.641 & 0.009 & 0.059 & 0.060 & 0.959 \\ 
    & $\mu_0(3)$  & 0.950 & 0.961 & 0.011 & 0.091 & 0.092 & 0.950 \\ 
   \hline
\multirow{4}{5.3em}{R-IPCW-Adm} & $\beta_1$   & 0.300 & 0.295 & -0.005 & 0.077 & 0.076 & 0.952 \\ 
   & $\beta_2$  & -0.300 & -0.296 & 0.004 & 0.149 & 0.148 & 0.954 \\ 
    & $\mu_0(1)$  & 0.632 & 0.640 & 0.008 & 0.063 & 0.064 & 0.962 \\ 
    & $\mu_0(3)$  & 0.950 & 0.958 & 0.008 & 0.101 & 0.103 & 0.954 \\ 
   \hline
\multirow{4}{5.3em}{R-S-IPCW-Adm} & $\beta_1$  & 0.300 & 0.296 & -0.004 & 0.076 & 0.076 & 0.954 \\ 
    & $\beta_2$ & -0.300 & -0.298 & 0.002 & 0.147 & 0.147 & 0.954 \\ 
    & $\mu_0(1)$  & 0.632 & 0.640 & 0.008 & 0.063 & 0.064 & 0.962 \\ 
    & $\mu_0(3)$  & 0.950 & 0.957 & 0.007 & 0.100 & 0.103 & 0.956 \\ 
   \hline
\end{tabular}
\caption{Simulation results for Ghosh-Lin model with independent censoring. 
	Sample size n=800, $\rho_1=1$ and $\rho_2=2$. 
	A-IPCW: only administrative
	censoring handled with IPCW adjustment; RA-IPCW: combined right
	censoring handled with IPCW; Adm: only administrative censoring
	handled with risk-set adjustment; R-IPCW-Adm: random censoring handled
	with IPCW weighting, and administrative censoring handled with risk-set
	adjustment;  R-IPCW-S-Adm: random censoring handled with stratified
	IPCW weighting, and administrative censoring handled with risk-set
	adjustment. Mean of estimates (Mean), bias of estimates (Bias),
	empirical standard error (EmpSD), mean of estimated standard errors
	(MeanSE), and coverage of 95 \% confidence intervals (Coverage)}
	\label{GL-12-indep}
\end{table}

%% file: GL-12-sim-A1.tex
\begin{table}[ht]
\centering
\begin{tabular}{l|l|rrrrrr}
  \hline
 Method & Par. & True & Mean & Bias & EmpSD & MeanSE & Coverage \\ 
 \hline
\multirow{4}{5.3em}{A-IPCW} & $\beta_1$   & 0.300 & 0.268 & -0.032 & 0.072 & 0.070 & 0.922 \\ 
  & $\beta_2$  & -0.300 & -0.273 & 0.027 & 0.139 & 0.138 & 0.946 \\ 
   & $\mu_0(1)$ & 0.632 & 0.640 & 0.008 & 0.061 & 0.061 & 0.946 \\ 
   & $\mu_0(3)$ & 0.950 & 0.960 & 0.010 & 0.093 & 0.094 & 0.954 \\ 
   \hline
\multirow{4}{5.3em}{RA-IPCW} & $\beta_1$   & 0.300 & 0.272 & -0.028 & 0.077 & 0.075 & 0.927 \\ 
  & $\beta_2$  & -0.300 & -0.276 & 0.024 & 0.148 & 0.146 & 0.952 \\ 
   & $\mu_0(1)$& 0.632 & 0.640 & 0.008 & 0.065 & 0.064 & 0.946 \\ 
   & $\mu_0(3)$ & 0.950 & 0.960 & 0.010 & 0.105 & 0.104 & 0.935 \\ 
   \hline
\multirow{4}{5.3em}{Adm} & $\beta_1$   & 0.300 & 0.298 & -0.002 & 0.073 & 0.071 & 0.937 \\ 
  & $\beta_2$  & -0.300 & -0.299 & 0.001 & 0.140 & 0.139 & 0.946 \\ 
    & $\mu_0(1)$ & 0.632 & 0.641 & 0.009 & 0.061 & 0.060 & 0.940 \\ 
   & $\mu_0(3)$ & 0.950 & 0.960 & 0.010 & 0.095 & 0.092 & 0.945 \\ 
   \hline
\multirow{4}{5.3em}{R-IPCW-Adm} & $\beta_1$   & 0.300 & 0.299 & -0.001 & 0.077 & 0.075 & 0.943 \\ 
  & $\beta_2$  & -0.300 & -0.299 & 0.001 & 0.148 & 0.147 & 0.956 \\ 
   & $\mu_0(1)$ & 0.632 & 0.641 & 0.009 & 0.066 & 0.064 & 0.943 \\ 
  & $\mu_0(3)$ & 0.950 & 0.961 & 0.011 & 0.107 & 0.104 & 0.928 \\ 
   \hline
\multirow{4}{5.3em}{R-S-IPCW-Adm} & $\beta_1$   & 0.300 & 0.299 & -0.001 & 0.076 & 0.075 & 0.943 \\ 
   & $\beta_2$  & -0.300 & -0.299 & 0.001 & 0.146 & 0.146 & 0.956 \\ 
   & $\mu_0(1)$ & 0.632 & 0.641 & 0.008 & 0.065 & 0.064 & 0.945 \\ 
    & $\mu_0(3)$ & 0.950 & 0.959 & 0.009 & 0.106 & 0.103 & 0.932 \\ 
   \hline
\end{tabular}
\caption{Simulation results for Ghosh-Lin model with dependent administrative censoring and independent random censoring. 
	Sample size n=800, $\rho_1=1$ and $\rho_2=2$. 
A-IPCW: only administrative censoring handled with IPCW adjustment; RA-IPCW: combined right censoring handled with IPCW; Adm: only administrative censoring handled with risk-set adjustment; R-IPCW-Adm: random censoring handled with IPCW weighting, and administrative censoring handled with risk-set adjustment;  R-IPCW-S-Adm: random censoring handled with stratified IPCW weighting, and administrative censoring handled with risk-set adjustment. Mean of estimates (Mean), bias of estimates (Bias), empirical standard error (EmpSD), mean of estimated standard errors (MeanSE), and coverage of 95 \% confidence intervals (Coverage)} 
\label{GL-12-depA}
\end{table}

%% file: GL-12-sim-A1-R1.tex
\begin{table}[ht]
\centering
\begin{tabular}{l|l|rrrrrr}
Method & Par.  & True & Mean & Bias & EmpSD & MeanSE & Coverage \\ 
\multirow{4}{5.3em}{A-IPCW} & $\beta_1$   & 0.300 & 0.267 & -0.033 & 0.072 & 0.070 & 0.917 \\ 
   & $\beta_2$  & -0.300 & -0.278 & 0.022 & 0.142 & 0.137 & 0.938 \\ 
  & $\mu_0(1)$ & 0.632 & 0.641 & 0.009 & 0.061 & 0.061 & 0.944 \\ 
  & $\mu_0(3)$ & 0.950 & 0.964 & 0.013 & 0.093 & 0.094 & 0.951 \\ 
   \hline
\multirow{4}{5.3em}{RA-IPCW} & $\beta_1$   & 0.300 & 0.242 & -0.058 & 0.076 & 0.074 & 0.873 \\ 
   & $\beta_2$ & -0.300 & -0.253 & 0.047 & 0.150 & 0.146 & 0.934 \\ 
  & $\mu_0(1)$ & 0.632 & 0.638 & 0.006 & 0.066 & 0.064 & 0.946 \\ 
  & $\mu_0(3)$ & 0.950 & 0.960 & 0.010 & 0.104 & 0.103 & 0.952 \\ 
   \hline
\multirow{4}{5.3em}{Adm} & $\beta_1$   & 0.300 & 0.296 & -0.004 & 0.072 & 0.071 & 0.943 \\ 
    & $\beta_2$  & -0.300 & -0.303 & -0.003 & 0.142 & 0.139 & 0.950 \\ 
   & $\mu_0(1)$ & 0.632 & 0.642 & 0.010 & 0.062 & 0.060 & 0.939 \\ 
   & $\mu_0(3)$ & 0.950 & 0.963 & 0.013 & 0.094 & 0.092 & 0.942 \\ 
   \hline
\multirow{4}{5.3em}{R-IPCW-Adm} & $\beta_1$   & 0.300 & 0.269 & -0.031 & 0.076 & 0.074 & 0.925 \\ 
   & $\beta_2$  & -0.300 & -0.278 & 0.022 & 0.151 & 0.147 & 0.947 \\ 
  & $\mu_0(1)$ & 0.632 & 0.641 & 0.008 & 0.066 & 0.065 & 0.946 \\ 
   & $\mu_0(3)$ & 0.950 & 0.962 & 0.012 & 0.105 & 0.104 & 0.949 \\ 
   \hline
\multirow{4}{5.3em}{R-S-IPCW-Adm} & $\beta_1$  & 0.300 & 0.296 & -0.004 & 0.076 & 0.074 & 0.947 \\ 
   & $\beta_2$  & -0.300 & -0.302 & -0.002 & 0.149 & 0.146 & 0.948 \\ 
   & $\mu_0(1)$ & 0.632 & 0.642 & 0.010 & 0.066 & 0.065 & 0.948 \\ 
   & $\mu_0(3)$ & 0.950 & 0.962 & 0.011 & 0.105 & 0.104 & 0.952 \\ 
   \hline
\end{tabular}
\caption{Simulation results for Ghosh-Lin model with dependent administrative censoring and dependent random censoring. 
	Sample size n=800, $\rho_1=1$ and $\rho_2=2$. 
	A-IPCW: only administrative censoring handled with IPCW adjustment; RA-IPCW: combined right censoring handeled with IPCW; Adm: only administrative censoring handled with risk-set adjustment; R-IPCW-Adm: random censoring handled with IPCW weighting, and administrative censoring handled with risk-set adjustment;  R-IPCW-S-Adm: random censoring handled with stratified IPCW weighting, and administrative censoring handled with risk-set adjustment. Mean of estimates (Mean), bias of estimates (Bias), empirical standard error (EmpSD), mean of estimated standard errors (MeanSE), and coverage of 95 \% confidence intervals (Coverage)} 
\label{GL-12-depAR}
\end{table}

%% file: GL-1Low-sim-indep.tex
\begin{table}[ht]
\centering
\begin{tabular}{l|l|rrrrrr}
   \hline
Method & Par.  & True & Mean & Bias & EmpSD & MeanSE & Coverage \\ 
   \hline
\multirow{4}{5.3em}{ A-IPCW} & $\beta_1$  & 0.300 & 0.300 & 0.000 & 0.060 & 0.060 & 0.944 \\ 
   & $\beta_2$ & -0.300 & -0.299 & 0.001 & 0.118 & 0.118 & 0.957 \\ 
   & $\mu_0(1)$  & 0.632 & 0.633 & 0.001 & 0.052 & 0.054 & 0.952 \\ 
  & $\mu_0(3)$  & 0.950 & 0.951 & 0.001 & 0.075 & 0.079 & 0.956 \\ 
   \hline
\multirow{4}{5.3em}{RA-IPCW} & $\beta_1$  & 0.300 & 0.301 & 0.001 & 0.067 & 0.065 & 0.941 \\ 
  & $\beta_2$ & -0.300 & -0.299 & 0.001 & 0.130 & 0.130 & 0.946 \\ 
   & $\mu_0(1)$  & 0.632 & 0.632 & 0.000 & 0.058 & 0.058 & 0.947 \\ 
   & $\mu_0(3)$  & 0.950 & 0.951 & 0.001 & 0.087 & 0.088 & 0.949 \\ 
   \hline
\multirow{4}{5.3em}{Adm} & $\beta_1$  & 0.300 & 0.301 & 0.001 & 0.061 & 0.060 & 0.944 \\ 
   & $\beta_2$ & -0.300 & -0.299 & 0.001 & 0.119 & 0.119 & 0.954 \\ 
   & $\mu_0(1)$  & 0.632 & 0.633 & 0.001 & 0.053 & 0.054 & 0.953 \\ 
  & $\mu_0(3)$ & 0.950 & 0.951 & 0.001 & 0.076 & 0.078 & 0.955 \\ 
   \hline
\multirow{4}{5.3em}{R-IPCW-Adm} & $\beta_1$  & 0.300 & 0.301 & 0.001 & 0.067 & 0.066 & 0.941 \\ 
   & $\beta_2$ & -0.300 & -0.299 & 0.001 & 0.131 & 0.130 & 0.946 \\ 
   & $\mu_0(1)$  & 0.632 & 0.632 & 0.000 & 0.059 & 0.058 & 0.945 \\ 
   & $\mu_0(3)$  & 0.950 & 0.951 & 0.001 & 0.087 & 0.088 & 0.947 \\ 
   \hline
\multirow{4}{5.3em}{R-S-IPCW-Adm} & $\beta_1$  & 0.300 & 0.302 & 0.002 & 0.067 & 0.066 & 0.942 \\ 
  & $\beta_1$ & -0.300 & -0.299 & 0.001 & 0.131 & 0.130 & 0.944 \\ 
   & $\mu_0(1)$  & 0.632 & 0.632 & -0.000 & 0.059 & 0.058 & 0.945 \\ 
   & $\mu_0(3)$  & 0.950 & 0.951 & 0.001 & 0.088 & 0.088 & 0.947 \\ 
   \hline
\end{tabular}
\caption{Simulation results for Ghosh-Lin model with independent censoring. 
	Sample size n=800, $\rho_1=1$ and $\rho_2=0.5$. 
	A-IPCW: only administrative censoring handled with IPCW adjustment; RA-IPCW: combined right censoring handled with IPCW; Adm: only administrative censoring handled with risk-set adjustment; R-IPCW-Adm: random censoring handled with IPCW weighting, and administrative censoring handled with risk-set adjustment;  R-IPCW-S-Adm: random censoring handled with stratified IPCW weighting, and administrative censoring handled with risk-set adjustment. Mean of estimates (Mean), bias of estimates (Bias), empirical standard error (EmpSD), mean of estimated standard errors (MeanSE), and coverage of 95 \% confidence intervals (Coverage)} 
	\label{GL-1Low-indep}
\end{table}

%% file: GL-1Low-sim-A1.tex
\begin{table}[ht]
\centering
\begin{tabular}{l|l|rrrrrr}
   \hline
Method & Par.  & True & Mean & Bias & EmpSD & MeanSE & Coverage \\ 
   \hline
\multirow{4}{5.3em}{ A-IPCW} & $\beta_1$  & 0.300 & 0.291 & -0.009 & 0.060 & 0.058 & 0.937 \\ 
  & $\beta_2$  & -0.300 & -0.287 & 0.013 & 0.121 & 0.117 & 0.940 \\ 
   & $\mu_0(1)$  & 0.632 & 0.631 & -0.001 & 0.053 & 0.054 & 0.942 \\ 
    & $\mu_0(3)$ & 0.950 & 0.948 & -0.003 & 0.078 & 0.078 & 0.943 \\ 
   \hline
\multirow{4}{5.3em}{RA-IPCW} & $\beta_1$  & 0.300 & 0.292 & -0.008 & 0.067 & 0.064 & 0.941 \\ 
  & $\beta_2$  & -0.300 & -0.287 & 0.013 & 0.135 & 0.129 & 0.935 \\ 
   & $\mu_0(1)$  & 0.632 & 0.630 & -0.002 & 0.058 & 0.058 & 0.946 \\ 
   & $\mu_0(3)$ & 0.950 & 0.946 & -0.004 & 0.088 & 0.087 & 0.946 \\ 
   \hline
\multirow{4}{5.3em}{Adm} & $\beta_1$  & 0.300 & 0.300 & 0.000 & 0.060 & 0.059 & 0.944 \\ 
   & $\beta_2$  & -0.300 & -0.296 & 0.004 & 0.122 & 0.117 & 0.944 \\ 
    & $\mu_0(1)$ & 0.632 & 0.632 & -0.000 & 0.053 & 0.053 & 0.942 \\ 
    & $\mu_0(3)$  & 0.950 & 0.949 & -0.001 & 0.079 & 0.078 & 0.941 \\ 
   \hline
\multirow{4}{5.3em}{R-IPCW-Adm} & $\beta_1$  & 0.300 & 0.301 & 0.001 & 0.067 & 0.065 & 0.940 \\ 
   & $\beta_2$  & -0.300 & -0.295 & 0.005 & 0.135 & 0.129 & 0.937 \\ 
   & $\mu_0(1)$  & 0.632 & 0.631 & -0.001 & 0.058 & 0.058 & 0.944 \\ 
   & $\mu_0(3)$  & 0.950 & 0.947 & -0.003 & 0.088 & 0.087 & 0.945 \\ 
   \hline
\multirow{4}{5.3em}{R-S-IPCW-Adm} & $\beta_1$  & 0.300 & 0.301 & 0.001 & 0.067 & 0.065 & 0.942 \\ 
   & $\beta_2$  & -0.300 & -0.295 & 0.005 & 0.135 & 0.129 & 0.936 \\ 
    & $\mu_0(1)$ & 0.632 & 0.631 & -0.001 & 0.058 & 0.058 & 0.946 \\ 
    & $\mu_0(3)$  & 0.950 & 0.946 & -0.004 & 0.088 & 0.087 & 0.944 \\ 
   \hline
\end{tabular}
\caption{Simulation results for Ghosh-Lin model with dependent administrative censoring and independent random censoring. Sample size n=800, $\rho_1=1$ and $\rho_2=0.5$. A-IPCW: only administrative censoring handled with IPCW adjustment; RA-IPCW: combined right censoring handled with IPCW; Adm: only administrative censoring handled with risk-set adjustment; R-IPCW-Adm: random censoring handled with IPCW weighting, and administrative censoring handled with risk-set adjustment;  R-IPCW-S-Adm: random censoring handled with stratified IPCW weighting, and administrative censoring handled with risk-set adjustment. Mean of estimates (Mean), bias of estimates (Bias), empirical standard error (EmpSD), mean of estimated standard errors (MeanSE), and coverage of 95 \% confidence intervals (Coverage)} 
	\label{GL-1Low-depA}
\end{table}

%% file: GL-1Low-sim-A1-R1.tex
\begin{table}[ht]
\centering
\begin{tabular}{l|l|rrrrrr}
  \hline
Method & Par.  & True & Mean & Bias & EmpSD & MeanSE & Coverage \\ 
  \hline
 \multirow{4}{5.3em}{ A-IPCW} & $\beta_1$   & 0.300 & 0.288 & -0.012 & 0.059 & 0.059 & 0.944 \\ 
  & $\beta_2$   & -0.300 & -0.293 & 0.007 & 0.118 & 0.117 & 0.946 \\ 
    & $\mu_0(1)$ & 0.632 & 0.634 & 0.002 & 0.055 & 0.054 & 0.945 \\ 
    & $\mu_0(3)$ & 0.950 & 0.951 & 0.001 & 0.079 & 0.079 & 0.945 \\ 
   \hline
\multirow{4}{5.3em}{RA-IPCW} & $\beta_1$  & 0.300 & 0.280 & -0.020 & 0.064 & 0.064 & 0.943 \\ 
  & $\beta_2$  & -0.300 & -0.284 & 0.016 & 0.127 & 0.127 & 0.956 \\ 
    & $\mu_0(1)$ & 0.632 & 0.632 & -0.001 & 0.061 & 0.058 & 0.934 \\ 
    & $\mu_0(3)$ & 0.950 & 0.947 & -0.003 & 0.090 & 0.087 & 0.935 \\ 
   \hline
\multirow{4}{5.3em}{Adm} & $\beta_1$   & 0.300 & 0.298 & -0.002 & 0.059 & 0.059 & 0.951 \\ 
  & $\beta_2$  & -0.300 & -0.302 & -0.002 & 0.119 & 0.117 & 0.948 \\ 
    & $\mu_0(1)$ & 0.632 & 0.635 & 0.003 & 0.055 & 0.054 & 0.940 \\ 
   & $\mu_0(3)$ & 0.950 & 0.952 & 0.002 & 0.080 & 0.078 & 0.942 \\ 
   \hline
\multirow{4}{5.3em}{R-IPCW-Adm} & $\beta_1$   & 0.300 & 0.288 & -0.012 & 0.064 & 0.064 & 0.949 \\ 
 & $\beta_2$  & -0.300 & -0.291 & 0.009 & 0.127 & 0.128 & 0.956 \\ 
    & $\mu_0(1)$ & 0.632 & 0.633 & 0.001 & 0.061 & 0.059 & 0.934 \\ 
   & $\mu_0(3)$ & 0.950 & 0.949 & -0.001 & 0.091 & 0.087 & 0.937 \\ 
   \hline
\multirow{4}{5.3em}{R-S-IPCW-Adm} & $\beta_1$   & 0.300 & 0.297 & -0.003 & 0.064 & 0.064 & 0.952 \\ 
   & $\beta_2$  & -0.300 & -0.299 & 0.001 & 0.127 & 0.127 & 0.953 \\ 
    & $\mu_0(1)$  & 0.632 & 0.634 & 0.002 & 0.061 & 0.059 & 0.937 \\ 
    & $\mu_0(3)$  & 0.950 & 0.950 & 0.000 & 0.091 & 0.088 & 0.938 \\ 
   \hline
\end{tabular}
\caption{Simulation results for Ghosh-Lin model with dependent administrative censoring and dependent random censoring. Sample size n=800 and $\rho_1=1$ and $\rho_2=0.5$. A-IPCW: only administrative censoring handled with IPCW adjustment, RA-IPCW: combined right censoring handled with IPCW, Adm: only administrative censoring handled with risk-set adjustment, R-IPCW-Adm: random censoring handled with IPCW weighting and administrative censoring handled with risk-set adjustment,  R-IPCW-S-Adm: random censoring handled with stratified IPCW weighting and administrative censoring handled with risk-set adjustment. Mean of estimates (Mean), bias of estimates (Bias), empirical standard error (EmpSD), mean of estimated standard errors (MeanSE), and coverage of 95 \% confidence intervals (Coverage)} 
	\label{GL-1Low-depAR}
\end{table}

%% file: FG-sim-indep.tex
\begin{table}[ht]
\centering
\begin{tabular}{l|l|rrrrrr}
   \hline
Method & Par. & True & Mean & Bias & EmpSD & MeanSE & Coverage \\ 
   \hline
\multirow{4}{5.3em}{RA-IPCW} & $\beta_1$   & 0.300 & 0.302 & 0.002 & 0.148 & 0.143 & 0.952 \\ 
  & $\beta_2$  & -0.300 & -0.301 & -0.001 & 0.279 & 0.278 & 0.951 \\ 
   & $\mu_0(1)$ & 0.085 & 0.084 & -0.001 & 0.020 & 0.020 & 0.956 \\ 
   & $\mu_0(3)$ & 0.243 & 0.243 & -0.000 & 0.046 & 0.050 & 0.972 \\ 
   \hline
\multirow{4}{5.3em}{Adm} & $\beta_1$   & 0.300 & 0.302 & 0.002 & 0.137 & 0.134 & 0.948 \\ 
  & $\beta_2$  & -0.300 & -0.297 & 0.003 & 0.260 & 0.260 & 0.954 \\ 
   & $\mu_0(1)$& 0.085 & 0.084 & -0.001 & 0.019 & 0.019 & 0.951 \\ 
   & $\mu_0(3)$ & 0.243 & 0.242 & -0.001 & 0.043 & 0.043 & 0.959 \\ 
   \hline
\multirow{4}{5.3em}{R-IPCW-Adm} & $\beta_1$   & 0.300 & 0.302 & 0.002 & 0.149 & 0.145 & 0.952 \\ 
   & $\beta_2$  & -0.300 & -0.301 & -0.001 & 0.280 & 0.280 & 0.952 \\ 
   & $\mu_0(1)$ & 0.085 & 0.084 & -0.001 & 0.020 & 0.020 & 0.953 \\ 
   & $\mu_0(3)$ & 0.243 & 0.243 & -0.000 & 0.047 & 0.047 & 0.961 \\ 
   \hline
\multirow{4}{5.3em}{R-S-IPCW-Adm} & $\beta_1$   & 0.300 & 0.305 & 0.005 & 0.149 & 0.144 & 0.949 \\ 
   & $\beta_2$  & -0.300 & -0.304 & -0.004 & 0.278 & 0.281 & 0.956 \\ 
    & $\mu_0(1)$ & 0.085 & 0.084 & -0.001 & 0.020 & 0.020 & 0.954 \\ 
    & $\mu_0(3)$ & 0.243 & 0.241 & -0.002 & 0.046 & 0.046 & 0.960 \\ 
  \end{tabular}
\caption{Simulation results for Fine-Gray model with independent censoring.
	RA-IPCW: combined right censoring handled with IPCW; Adm: only administrative censoring handled with risk-set adjustment; R-IPCW-Adm: random censoring handled with IPCW weighting, and administrative censoring handled with risk-set adjustment;  R-S-IPCW-Adm: random censoring handled with stratified IPCW weighting, and administrative censoring handled with risk-set adjustment. Mean of estimates (Mean), bias of estimates (Bias), empirical standard error (EmpSD), mean of estimated standard errors (MeanSE), and coverage of 95 \% confidence intervals (Coverage)} 
	\label{FG-indep}
\end{table}

%% file: FG-sim-A1.tex
\begin{table}[ht]
\centering
\begin{tabular}{l|l|rrrrrr}
Method & Par.  & True & Mean & Bias & EmpSD & MeanSE & Coverage \\ 
 \multirow{4}{5.3em}{RA-IPCW} & $\beta_1$   & 0.300 & 0.274 & -0.026 & 0.139 & 0.140 & 0.949 \\ 
  & $\beta_2$  & -0.300 & -0.280 & 0.020 & 0.278 & 0.276 & 0.953 \\ 
   & $\mu_0(1)$  & 0.085 & 0.084 & -0.001 & 0.020 & 0.020 & 0.957 \\ 
    & $\mu_0(3)$  & 0.243 & 0.243 & 0.000 & 0.046 & 0.049 & 0.967 \\ 
   \hline
\multirow{4}{5.3em}{Adm} & $\beta_1$   & 0.300 & 0.307 & 0.007 & 0.129 & 0.131 & 0.958 \\ 
   & $\beta_2$  & -0.300 & -0.304 & -0.004 & 0.259 & 0.257 & 0.951 \\ 
   & $\mu_0(1)$  & 0.085 & 0.084 & -0.001 & 0.019 & 0.019 & 0.955 \\ 
   & $\mu_0(3)$  & 0.243 & 0.243 & -0.001 & 0.043 & 0.043 & 0.956 \\ 
   \hline
\multirow{4}{5.3em}{R-IPCW-Adm} & $\beta_1$   & 0.300 & 0.307 & 0.007 & 0.140 & 0.141 & 0.959 \\ 
   & $\beta_2$  & -0.300 & -0.309 & -0.009 & 0.280 & 0.278 & 0.952 \\ 
    & $\mu_0(1)$  & 0.085 & 0.084 & -0.001 & 0.020 & 0.020 & 0.955 \\ 
    & $\mu_0(3)$  & 0.243 & 0.243 & -0.000 & 0.047 & 0.047 & 0.953 \\ 
   \hline
\multirow{4}{5.3em}{R-S-IPCW-Adm} & $\beta_1$   & 0.300 & 0.309 & 0.009 & 0.140 & 0.141 & 0.959 \\ 
   & $\beta_2$  & -0.300 & -0.309 & -0.009 & 0.277 & 0.278 & 0.956 \\ 
   & $\mu_0(1)$  & 0.085 & 0.084 & -0.001 & 0.020 & 0.020 & 0.953 \\ 
   & $\mu_0(3)$  & 0.243 & 0.241 & -0.002 & 0.046 & 0.046 & 0.952 \\ 
  \end{tabular}
\caption{Simulation results for Fine-Gray model with dependent administrative censoring and independent random censoring. 
	RA-IPCW: combined right censoring handeled with IPCW; Adm: only administrative censoring handled with risk-set adjustment; R-IPCW-Adm: random censoring handled with IPCW weighting, and administraive censoring handled with risk-set adjustment;  R-S-IPCW-Adm: random censoring handled with stratified IPCW weighting, and administraive censoring handled with risk-set adjustment. Mean of estimates (Mean), bias of estimates (Bias), empirical standard error (EmpSD), mean of estimated standard errors (MeanSE), and coverage of 95 \% confidence intervals (Coverage)} 
	\label{FG-depA}
\end{table}

%% file: FG-sim-A1-R1.tex
\begin{table}[ht]
\centering
\begin{tabular}{l|l|rrrrrr}
   \hline
Method & Par.  & True & Mean & Bias & EmpSD & MeanSE & Coverage \\ 
   \hline
 \multirow{4}{5.3em}{RA-IPCW} & $\beta_1$  & 0.300 & 0.238 & -0.062 & 0.150 & 0.145 & 0.920 \\ 
  & $\beta_2$  & -0.300 & -0.251 & 0.049 & 0.290 & 0.290 & 0.957 \\ 
   & $\mu_0(1)$  & 0.085 & 0.083 & -0.002 & 0.021 & 0.021 & 0.957 \\ 
   & $\mu_0(3)$ & 0.243 & 0.240 & -0.003 & 0.048 & 0.052 & 0.969 \\ 
   \hline
 \multirow{4}{5.3em}{Adm} & $\beta_1$  & 0.300 & 0.305 & 0.005 & 0.134 & 0.131 & 0.947 \\ 
  & $\beta_2$ & -0.300 & -0.307 & -0.007 & 0.265 & 0.258 & 0.944 \\ 
    & $\mu_0(1)$  & 0.085 & 0.084 & -0.001 & 0.019 & 0.019 & 0.948 \\ 
   & $\mu_0(3)$ & 0.243 & 0.242 & -0.002 & 0.044 & 0.043 & 0.949 \\ 
   \hline
 \multirow{4}{5.3em}{R-IPCW-Adm} & $\beta_1$  & 0.300 & 0.270 & -0.030 & 0.152 & 0.147 & 0.940 \\ 
  & $\beta_2$ & -0.300 & -0.279 & 0.021 & 0.292 & 0.292 & 0.959 \\ 
   & $\mu_0(1)$ & 0.085 & 0.084 & -0.001 & 0.021 & 0.021 & 0.953 \\ 
   & $\mu_0(3)$ & 0.243 & 0.241 & -0.003 & 0.049 & 0.049 & 0.957 \\ 
   \hline
 \multirow{4}{5.3em}{R-S-IPCW-Adm} & $\beta_1$  & 0.300 & 0.303 & 0.003 & 0.152 & 0.147 & 0.944 \\ 
  & $\beta_2$  & -0.300 & -0.309 & -0.009 & 0.291 & 0.290 & 0.956 \\ 
   & $\mu_0(1)$  & 0.085 & 0.084 & -0.001 & 0.021 & 0.021 & 0.950 \\ 
    & $\mu_0(3)$ & 0.243 & 0.240 & -0.004 & 0.049 & 0.049 & 0.957 \\ 
  \end{tabular}
\caption{Simulation results for Fine-Gray model with dependent administrative censoring and dpendent random censoring. 
	RA-IPCW: combined right censoring handeled with IPCW; Adm: only
	administrative censoring handled with risk-set adjustment; R-IPCW-Adm:
	random censoring handled with IPCW weighting, and administraive
	censoring handled with risk-set adjustment;  R-S-IPCW-Adm: random
	censoring handled with stratified IPCW weighting, and administraive
	censoring handled with risk-set adjustment. Mean of estimates (Mean),
	bias of estimates (Bias), empirical standard error (EmpSD), mean of
	estimated standard errors (MeanSE), and coverage of 95 \% confidence
	intervals (Coverage)} 
	\label{FG-depAR}
\end{table}